\documentstyle[12pt,epsfig]{article}
\newlength{\dinwidth}
\newlength{\dinmargin}
\newcommand\llambda{\zeta}
\newcommand\ben{\begin{equation}}
\newcommand\een{\end{equation}}
\newcommand\bea{\begin{eqnarray}}
\newcommand\eea{\end{eqnarray}}

\newcommand\nn{\nonumber}

\newcommand{\beq}{\begin{equation}}
\newcommand{\eeq}{\end{equation}}
\newcommand{\beqa}{\begin{eqnarray}}
\newcommand{\eeqa}{\end{eqnarray}}
\newcommand{\be}{\begin{equation}}
\newcommand{\ee}{\end{equation}}
\newcommand{\bc}{\begin{center}}
\newcommand{\ec}{\end{center}}

\newcommand{\cO}{{\mathcal O}}

\newcommand\hcOP{\hat{\mathcal{O}}^\Phi}

\begin{document}
\thispagestyle{empty}
\addtocounter{page}{-1}
\vskip-0.35cm
\begin{flushright}
UK/09-03, SITP/09-28, SLAC-PUB-13683,TIFR-TH/09-16 \\
\end{flushright}
\vspace*{0.2cm}
\centerline{\Large \bf   Slowly Varying Dilaton Cosmologies}
\vspace*{0.1cm}
\centerline{\Large \bf and their Field Theory  Duals}
\vspace*{0.5cm}
\centerline{\bf Adel Awad${}^{a,b}$
Sumit R. Das${}^c$, Archisman Ghosh ${}^c$}
\centerline{\bf Jae-Hyuk Oh ${}^c$,
Sandip P. Trivedi ${}^{d,e,f}$}
\vspace*{0.4cm}
\centerline{\it Center for Theoretical Physics, British University of
  Egypt}
\vspace*{0.1cm}
\centerline{\it Sherouk City 11837, P.O. Box 43, EGYPT ${}^a$}
\vspace*{0.2cm}
\centerline{\it Department of Physics, Faculty of Science,}
\vspace*{0.1cm}
\centerline{\it Ain Shams University, Cairo, 11566, EGYPT ${}^b$}
\vspace*{0.2cm}
\centerline{\it Department of Physics and Astronomy,}
\vspace*{0.1cm}
\centerline{\it University of Kentucky, Lexington, KY 40506 \rm USA ${}^c$}
\vspace*{0.2cm}
\centerline{\it Tata Institute of Fundamental Research, Mumbai 400005, \rm INDIA ${}^d$}
\vspace*{0.2cm}
\centerline{\it Stanford Institute of Theoretical Physics, Stanford CA 94305 USA ${}^e$}
\vspace{0.2cm}
\centerline{\it SLAC, Stanford University, Stanford, CA 94309 USA ${}^f$}
\vspace{0.5cm}
\centerline{\tt adel@pa.uky.edu, das@pa.uky.edu, archisman.ghosh@uky.edu}
\centerline{\tt jaehyukoh@uky.edu, trivedi.sp@gmail.com}

\vspace*{0.5cm} \centerline{\bf Abstract} \vspace*{0.3cm}
We consider a deformation of the $AdS_5\times S^5$ solution of IIB
supergravity obtained by taking the boundary value of the dilaton to
be time dependent.  The time dependence is taken to be slowly varying
on the AdS scale thereby introducing a small parameter $\epsilon$. The
boundary dilaton has a profile which asymptotes to a constant in the
far past and future and attains a minimum value at intermediate times.
We construct the sugra solution to first non-trivial order in
$\epsilon$, and find that it is smooth, horizon free, and
asymptotically $AdS_5\times S^5$ in the far future. When the
intermediate values of the dilaton becomes small enough the curvature
becomes of order the string scale and the sugra approximation breaks
down. The resulting dynamics is analysed in the dual $SU(N)$ gauge
theory on $S^3$ with a time dependent coupling constant which varies
slowly.  When $N \epsilon \ll 1$, we find that a quantum adiabatic
approximation is applicable, and use it to argue that at late times
the geometry becomes smooth $AdS_5\times S^5$ again. When $N \epsilon
\gg 1$, we formulate a classical adiabatic perturbation theory based
on coherent states which arises in the large $N$ limit. For large
values of the 'tHooft coupling this reproduces the supergravity
results. For small 'tHooft coupling the coherent state calculations
become involved and we cannot reach a definite conclusion.  We argue
that the final state should have a dual description which is mostly
smooth $AdS_5$ space with the possible presence of a small black hole.

\vspace*{0.5cm}
\baselineskip=18pt


\newpage

\section{Introduction}

The AdS/CFT correspondence \cite{Maldacena1, GKP, Witten1} provides us
with a non-perturbative formulation of quantum gravity. One hopes that
it will shed some light on the deep mysteries of quantum gravity, in
particular on the question of singularity resolution.

Motivated by this hope we consider a class of time dependent solutions
in this paper which can be viewed as deformations of the $AdS_5 \times
S^5$ background in IIB string theory. These solutions are obtained by
taking the boundary value of the dilaton in AdS space to become time
dependent \footnote{It is important in the subsequent discussion that we
work in global $AdS_5$ with the boundary $S^3 \times R$.} . We are
free to take the boundary value of the dilaton to be any time
dependent function. To keep the solutions under analytical control
though we take the rate of  time variation of the dilaton to be
small compared to the radius of AdS space, $R_{AdS}$. This introduces
a small parameter $\epsilon$ and we construct the bulk solution in
perturbation theory in $\epsilon$.  The resulting solutions are found
to be well behaved.  In particular one finds that no black hole
horizon forms in the course of time evolution. The metric and dilaton
respond on a time scale of order $R_{AdS}$ which is nearly
instantaneous compared to the much slower time scale at which the
boundary value of the dilaton varies. For dilaton profiles which
asymptote to a constant in the far future one finds that all the
energy that is sent in comes back out and the geometry settles down
eventually to that of $AdS$ space.  What makes these solutions
non-trivial is that by waiting for a long enough time, of order
$R_{AdS}\over \epsilon$, a big change in the boundary dilaton can
occur. The solutions probe the response of the bulk to such big
changes.

Consider an example of this type where the boundary dilaton undergoes
a big change making the 'tHooft coupling\footnote{When we refer to the
'tHooft coupling we have the gauge theory in mind and accordingly by
the dilaton in this context we will always mean its boundary value.}
of order unity or smaller at intermediate times,
\be
\label{thooft}
\lambda \equiv g_s N \le O(1),
\ee
when
\footnote{Here $N$ is the number of units of flux in the bulk and
the rank of the gauge group in the boundary theory.} $t\simeq 0$,
before becoming large again in the far future. As was mentioned above,
the bulk responds rapidly to the changing boundary conditions and
within a time of order $R_{AdS}$ the dilaton everywhere in the bulk
then becomes small and meets the condition, eq.(\ref{thooft}).  Now
the supergravity solution receives $\alpha'$ corrections in string
theory, these are important when $R_{AdS}$ becomes of order the string
scale. Using the well known relation,
\be
\label{radiusa}
R_{AdS}/l_s \sim (g_s N)^{1\over 4}
\ee
we then find that once eq.(\ref{thooft}) is met the curvature becomes
of order the string scale everywhere along a space-like slice which
intersects the boundary.  As a result the supergravity approximation
breaks down along this slice and the higher derivative corrections
becomes important for the subsequent time development. This break down
of the supergravity approximation is the sense in which a singularity
arises in these solutions.

In contrast the curvature in units of the 10-dim. Planck scale
$l_{Pl}$ (or the $5$-dim Planck scale) remains small for all time. The
radius $R_{AdS}$ in $l_{Pl}$ units is given by,
\be
\label{radiusb}
R_{AdS}/l_{Pl} \sim N^{1\over 4}
\ee
We keep $N$ to be fixed and large throughout the evolution, this then
keeps the curvature small in Planck units \footnote{The backreaction
corrects the curvature but these corrections are suppressed in
$\epsilon$.}. The solutions we consider can therefore be viewed in the
following manner: the curvature in Planck units in these solutions
stays small for all time, but for a dilaton profile which meets the
condition eq.(\ref{thooft}) the string scale in length grows and
becomes of order the curvature scale at intermediate times. At this
stage the geometry gets highly curved on the string scale. We are
interested in whether a smooth spacetime geometry can emerge again in
the future in such situations.

It is worth relating this difference in the behaviour of the
curvature as measured in  string and Planck scales to
another fact. We saw that when the curvature becomes of order the
string scale $\alpha'$ corrections become important. The second source
of corrections to the supergravity approximation are quantum loop
corrections.  Their importance is determined by the parameter
$1/N$. Since $N$ is kept fixed and large these corrections are always
small. From eq.(\ref{radiusb}) we see that this  ties into the fact
that the AdS radius stays large in Planck units.

To understand the evolution of the system once the curvature gets to
be of order the string scale we turn to the dual gauge theory. The
gauge theory lives on an $S^3$ of radius $R$ and the slowly varying
dilaton maps to a Yang-Mills coupling which varies slowly compared to
$R$. Since these are the only two length scales in the system the slow
time variation suggests that one can understand the resulting dynamics
in terms of an adiabatic approximation.

In fact we find it useful to consider two different adiabatic
perturbation theories. The first, which we call quantum adiabatic
perturbation theory is a good approximation when the parameter
$\epsilon$ satisfies the condition,
\be
\label{condall}
N \epsilon \ll 1.
\ee
Once this condition is met the rate of change of the Hamiltonian is
much smaller than the energy gap between the ground state and the
first excited state in the gauge theory. As a result the standard text
book adiabatic approximation in quantum mechanics applies and the
system at any time is, to good approximation, in the ground state of
the {\it instantaneous} Hamiltonian. In the far future, when the time
dependence turns off, the state settles into the ground state of the
resulting ${\cal N}=4$ SYM theory, and admits a dual description as a
smooth AdS space.

Note that this argument holds even when the 'tHooft coupling at
intermediate times becomes of order unity or smaller. The fact that
the states of the time independent ${\cal N}=4$ SYM theory furnish a
unitary representation of the conformal group guarantees that the
spectrum has a gap of order $1/R$ for all values of the Yang Mills
coupling, \cite{Mack}, see also, \cite{Shiraz0}, \cite{Magoo}.
Thus as long as eq.(\ref{condall})
is met the conditions for this perturbation theory apply. As a result,
we learn that for very slowly varying dilaton profiles which meet the
condition, eq.(\ref{condall}), the geometry after becoming of order
the string scale at intermediate times, again opens out into a smooth
$AdS$ space in the far future.

The supergravity solutions we construct are controlled in the approximation,
\be
\label{condee}
\epsilon \ll 1.
\ee
This is different, and much less restrictive,
than the condition stated above in eq.(\ref{condall}) for the validity
of the quantum adiabatic perturbation theory.  In fact one finds that
a different perturbation theory can also be formulated in the gauge
theory. This applies when the conditions,
\be
\label{condcap1}
N\epsilon \gg 1, \ \  \epsilon \ll 1
\ee
are met. This approximation is classical in nature and arises because
the system is in the large N-limit (otherwise eq.(\ref{condcap1})
cannot be met).  We will call this approximation the ``Large N
Classical Adiabatic Perturbation Theory'' (LNCAPT) below. The
behaviour of the system in this approximation reproduces the behaviour
of the supergravity solutions for cases where the 'tHooft coupling is
large for all times.

Let us now discuss this approximation in more detail.
Each gauge invariant operator in the boundary theory gives rise to an infinite
 tower of coupled oscillators whose frequency grows with growing mode
 number. The gauge invariant operators are dual to bulk modes. The
 infinite tower of oscillators which arises for each operator is dual
 to the infinite number of modes, with different radial wave functions
 and different frequency, which arise for each bulk field. Of
 particular importance is the operator dual to the dilaton $\hat{{\cal
 O}}$ and the modes which arise from it. The time varying
 boundary dilaton results in a driving force for these oscillators.
 When $N \epsilon \gg 1$, these oscillators are excited by the driving
 force into a coherent state with a large mean occupation number of
 quanta, of order $N \epsilon$, and therefore behave classically.
This is a reflection of the fact that at large $N$, the system
behaves classically : coherent states of these oscillators
correspond to classical configurations (see e.g. Ref
 \cite{Yaffe:1981vf}).

Usually a reformulation of the boundary theory in terms of such
oscillators is not very useful, since these oscillators would have a
nontrivial operator algebra which would signify that the bulk modes
are interacting. Simplifications happen
in low dimensional situations
like Matrix Quantum Mechanics \cite{Jevicki:1979mb}
where one is led to a collective field
theory in $1+1$ dimensions  as an explicit construction
of the holographic map \cite{Das:1990kaa}.
Even in this situation, the collective field
theory is a nontrivial interacting theory, i.e. the oscillators are
coupled. In our case there are an infinite number of collective fields
which would seem to make the situation hopeless.

In our setup, however, the slowness of the driving force simplifies the
situation drastically.
The source couples directly to the dilaton in the bulk, and
when $\epsilon \ll 1$, to lowest order the response of the dilaton as
well as the other fields is {\em linear} and independent of each
other. This will be clear in the supergravity solutions we present
below. This implies that to lowest order in $\epsilon$, the
oscillators which are dual to  these modes are really {\em harmonic
oscillators} which are decoupled from each other.

The resulting dynamics is then well approximated by the classical
 adiabatic perturbation theory, which we refer to as the LNCAPT as
 mentioned above. The criterion for its applicability is that the
 driving force varies on a time scale much slower than the frequency
 of each oscillator. In particular if the frequency of the driving
 force is of order that of the oscillators one would be close to
 resonance and the perturbation theory would break down. In our case
 this condition for the driving force to vary slowly compared to the
 frequency of the oscillators, becomes eq.(\ref{condee}). When this
 condition is met, the adiabatic approximation is valid for all modes
 - even those with the lowest frequency.  The expectation value of the
 energy and the  operator dual to the dilaton, ${\hat{\mathcal{O}}}$, can then be
 calculated in the resulting perturbation theory and we find that the
 leading order answers in $\epsilon$ agree with the supergravity
 calculations \footnote{More precisely, both the supergravity and the
 forced oscillator calculations need to be renormalised to get finite
 answer. One finds that after the counter terms are chosen to get
 agreement for the standard two point function ( which measures the
 response for a small amplitude dilaton perturbation) the expectation
 value of the energy and ${\hat O}$, agree.}.

Having understood the supergravity solutions in the gauge theory
  language we turn to asking what happens if the 'tHooft coupling
  becomes of order unity or smaller at intermediate times (while still
  staying in the parametric regime eq.(\ref{condcap1})).  The new
  complication is that additional oscillators now enter the analysis.
  These oscillators correspond to string modes in the bulk. When the
  'tHooft coupling becomes of order unity their frequencies can become
  small and comparable to the oscillators which are dual to supergravity
  modes.

At first sight one is tempted to conclude that these additional
  oscillators do not change the dynamics in any significant manner and
  the system continues to be well approximated by the large N
  classical adiabatic approximation. The following arguments support
  this conclusion. First, the anharmonic terms continue to be of order
  $\epsilon$ and thus are small, so that the  oscillators are
  approximately decoupled. Second, the existence of a gap of order
  $1/R$ for all values of the 'tHooft
  coupling, which we referred to above,  ensures that the driving force
  varies much more slowly than
  the frequency of the additional oscillators, thus keeping the
  system far from resonance. Finally, one still expects that in the
  parametric regime, eq.(\ref{condcap1}),  an  $O(N \epsilon)$ number
  of quanta are  produced keeping the system classical. These
  arguments suggest that the system should continue to be well
  approximated by the LNCAPT. In fact, since the additional
  oscillators do not directly couple to the driving force produced by
  the time dependent dilaton, but rather couple to it only through
  anharmonic terms which are subdominant in $\epsilon$, their effects
  should be well controlled in an $\epsilon$ expansion. If these
  arguments are correct the energy which is pumped into the system
  initially should then get completely pumped back out and the system
  should settle into the ground state of the final ${\cal N}=4$ theory
  in the far future. The dual description in the far future would then
  be a smooth $AdS_5$ space-time.

However, further thought suggests another possibility for the
resulting dynamics which is of a qualitatively different kind. This
possibility arises because, as was mentioned above, when the 'tHooft
coupling becomes of order unity string modes can get as light as
supergravity modes. This means that the frequency of some of the
oscillators dual to string modes can become comparable to oscillators
dual to supergravity modes, and thus the string mode oscillators can
get activated. Now there are many more string mode oscillators than
there are supergravity mode oscillators, since the supergravity modes correspond to
chiral operators in the gauge theory which are only $O(1)$ in number,
while the string modes correspond to non-chiral operators which are
$O(N^2)$ in number. Thus once string mode oscillators can get
activated there is the possibility that many new degrees of freedom
enter the dynamics.

With so many degrees of freedom available the system could thermalise
at least in the large $N$ limit. In this case the energy which is
initially present in the oscillators that directly couple to the
dilaton would get equi-partitioned among all the degrees of
freedom. The subsequent evolution would be dissipative and this energy
would not be recovered in the far future. At late times, when the
'tHooft coupling becomes big again, the gravity description of the
dissipative behaviour depends on how small is $\epsilon$.  From the
calculations done in the supergravity regime one knows that the total
energy that is produced is of order $N^2\epsilon^2$. When $N\epsilon
\gg 1$, but $\epsilon \ll (g_{YM}^2N)^{-7/8}$ the result is likely to
be a gas of string modes.  However if  $\epsilon > (g_{YM}^2N)^{-7/8}$,
the energy is sufficient to form a small black hole (with horizon
radius smaller than $R_{AdS}$).  A big black hole cannot form since
this would require an energy of the order of $N^2$,  and $\epsilon \ll
1$ always.  Thus, in the far future, once the 'tHooft coupling becomes
large again, the strongest departure from normal space-time would be
the presence of a small black hole in AdS space.  The small black hole
would eventually disappear by emitting Hawking radiation but that
would happen on a much longer time scale of order $N^2 R_{AdS}$.

It is difficult for us to settle here which of the two possibilities
discussed above, either adiabatic non-dissipative behaviour well
described by the LNCAPT, or dissipative behaviour with organised
energy being lost in heat, is the correct one.  One complication is
that the rate of time variation which is set by $\epsilon$ is also the
strength of the anharmonic couplings between the oscillators. In
thermodynamics, working in the microcanonical ensemble, it is well
known that with energy of order $N^2 \epsilon^2$ the configuration
which entropically dominates is a small black hole \footnote{At least
when the 'tHooft coupling is big enough so that supergravity can be
trusted.}. This suggests that if the time variation in the problem
were much smaller than the anharmonic terms a small black hole would
form. However, in our case their being comparable makes it a more
difficult question to decide.  One should emphasise that regardless of
which possibility is borne out our conclusion is that most of the
space time in the far future is smooth AdS, with the possible presence
of a small black hole.

Let us end with some comments on related work.  The spirit of our
investigation is close to the work on AdS cosmologies in
\cite{adnt} and related work in
\cite{Das1} - \cite{Madhu:2009jh}.  See also
\cite{Chen:2005ae}, \cite{Lin:2006ie}, \cite{ohta} for additional
work. Discussion of cosmological singularities in the context of
Matrix Theory appears in \cite{Craps1}.

The supergravity analysis we describe is closely related to the
strategy which was used in the paper \cite{BLMNTW}, for finding forced
fluid dynamics solutions; in that case one worked with an infinite
brane at temperature $T$ and the small parameter was the rate of
variation of the dilaton (or metric) compared to $T$. Our regime of
interest is complementary to that in \cite{Bhattacharyya:2009uu} where
the dilaton was chosen to be small in amplitude, but with arbitrary
time dependence and which leads to formation of black holes in
supergravity for a suitable regime of parameters.

This paper is organised as follows. In section \S2 we find the
supergravity solutions and use them to find the expectation value of
operators in the boundary theory like the stress energy and
$\hat{{\cal O}}$ in \S3.  The quantum adiabatic perturbation theory is
discussed in \S4. A forced harmonic oscillator is discussed in
\S5. This simple system helps illustrate the difference between the
two kinds of perturbation theory and sets the stage for the discussion
of the The Large N classical adiabatic approximation in
\S6. Conclusions and future directions are discussed in \S7.  There
are three appendices which contains details of derivation of some of
the formulae in the main text.

\section{The Bulk Response}

In this section we will calculate the deformation of the supergravity
solution in the presence of a slowly varying time
dependent but spatially homogeneous dilaton specified on the boundary.
This will be a reliable description of the time evolution of the
system so long as $e^{\Phi(t)}$ never becomes small.

\subsection{Some General Considerations}

IIB supergravity in the presence of the RR five form flux is well
known to have an $AdS_5\times S^5$ solution. In global coordinates this
takes the form,
\be
\label{metgc}
ds^2=-(1+{r^2\over R_{AdS}^2})dt^2 +{dr^2\over 1+{r^2\over R_{AdS}^2}}
+ r^2d\Omega_3^2 +R_{AdS}^2 d\Omega_5^2.
\ee
Here $R_{AdS}$ is given by,
\be
\label{valrads}
R_{AdS}=(4\pi g_s N)^{1/4}l_s \sim N^{1/4} l_{pl}
\ee
where $l_s$ is
the string scale and $l_{pl}\sim g_s^{1/4} l_s$ is the ten dimensional
Planck scale.  $g_s$ is the value of the dilaton, which is constant
and does not vary with time or spatial position,
\be
\label{dil0}
e^\Phi=g_s.
\ee

In the time dependent situations we consider below $N$ will be held fixed.
Let us discuss some of our convention es before proceeding. We will find it convenient to work in the $10$-dim. Einstein frame.
Usually one fixes $l_{Pl}$ to be of order unity in this frame. Instead for our purposes it will be convenient to set
\be
 \label{radsunity}
 R_{AdS}=1.
\ee
From eq.(\ref{valrads}) this means setting $l_{Pl} \sim 1/N^{1/4}$.
The $AdS_5\times S^5$ solution then becomes,
\be
\label{zzmetric}
ds^2=-(1+r^2)dt^2+{1\over (1+r^2)} dr^2 + r^2 d\Omega_3^2+ d\Omega_5^2,
\ee
for any constant value of the dilaton, eq.(\ref{dil0}).
Let us also mention that when we turn to the boundary gauge theory we will set the
radius $R$ of the $S^3$ on which it lives to also be unity.

 The essential idea in finding the solutions we describe is the
following. Consider a situation where $\Phi$ varies with time slowly
compared to $R_{AdS}$.  Since the solution above exists for any value
of $g_s$ and the dilaton varies slowly one expects that the resulting
metric at any time $t$ is well approximated by the $AdS_5\times S^5$
metric  given in eq.(\ref{zzmetric}).  This zeroth order metric will be corrected due to the varying
dilaton which provides an additional source of stress energy in the
Einstein equations.  However these changes should be small for a
slowly varying dilaton and should therefore be calculable order by
order in perturbation theory.

Let us make this more precise.  Consider as the starting point of this
perturbation theory the $AdS_5$ metric given in eq.(\ref{zzmetric}) and a
dilaton profile,
\be
\label{dilprof}
\Phi=\Phi_0(t)
\ee
which is a function of time alone. We take $\Phi_0(t)$ to be of the form,
\be
\label{dimpara}
\Phi_0 = f({\epsilon t\over R_{AdS}}) \ee
where $f({\epsilon t\over
R_{AdS}})$ is dimensionless function of time and $\epsilon$ is a small
parameter,
\be
\label{conde}
\epsilon \ll 1.
\ee
The function $f$ satisfies the property that
\be
\label{dervf}
f'({\epsilon t \over R_{AdS}})\sim O(1)
\ee
where prime indicates derivative with respect to the argument of $f$.

 When $\epsilon=0$, the dilaton is a constant and the solution reduces
to $AdS_5\times S^5$. When $\epsilon$ is small,
\be
\label{dervdil}
{d\Phi_0\over dt}={\epsilon \over R_{AdS}} f'({\epsilon t \over
R_{AdS}})\sim {\epsilon \over R_{AdS}}
\ee
so that the dilaton is
varying slowly on the scale $R_{AdS}$, and the contribution that the
dilaton makes to the stress tensor is parametrically suppressed \footnote{The more
precise statement for the slowly varying nature of the dilaton,
as will be discussed in a footnote before eq.(\ref{oscconda}), is that its
 Fourier transform has support at frequencies much smaller than $1/R_{AdS}$.}. In
such a situation the back reaction can be calculated order by order in
$\epsilon$.  The time dependent solutions we consider will be of this
type and $\epsilon$ will play the role of the small parameter in which
we carry out the perturbation theory.  A simple rule to count powers
of $\epsilon$ is that every time derivative of $\Phi_0$ comes with a
factor of $\epsilon$.

The profile for the dilaton we have considered in eq.(\ref{dilprof}) is $S^5$ symmetric. It is consistent  to assume that the back reacted metric
will also be $S^5$ symmetric with the radius of the $S^5$ being equal to
$R_{AdS}$.  The interesting time dependence will then unfold in the
remaining five directions of $AdS$ space and we will focus on them in
the following analysis.

The zeroth order metric in these directions  is given by,
\be
\label{zmetric}
ds^2=-(1+r^2)dt^2+{1\over (1+r^2)} dr^2 + r^2 d\Omega_3^3.
\ee
And the zeroth order dilaton is given by eq.(\ref{dilprof}),
\be
\label{redil}
\Phi_0=f(\epsilon t).
\ee
We can now calculate the corrections to
this solution order by order in $\epsilon$.

Let us make two more points at this stage.  First, we will consider a
dilaton profile $\Phi_0$ which approaches a constant as $t\rightarrow
-\infty$. This means that in the far past the corrections to the
metric and the dilaton which arise as a response to the time variation
of the dilaton must also vanish. Second, the perturbation theory we
have described above is a derivative expansion. The solutions we find
can only describe slowly varying situations. This stills allows for a
big change in the amplitude of the dilaton and the metric though, as
long as such changes accrue gradually. It is this fact that makes the
solutions non-trivial.

\subsection{Corrections to the Dilaton}

Let us first calculate the corrections to the dilaton.  We can expand the
dilaton as,
\be
\label{exdil}
\Phi(t)=\Phi_0(t) + \Phi_1(r,t) + \Phi_2(r,t) \cdots,
\ee
where
$\Phi_0$ is the zeroth order profile we start with, given in
eq.(\ref{dimpara}). $\Phi_1$ is of order $\epsilon$, $\Phi_2$ is of
order $\epsilon^2$ and so on.  The metric can be expanded as,
\be
\label{exmetric}
g_{ab}=g^{(0)}_{ab} +  g^{(1)}_{ab}+  g^{(2)}_{ab}+\cdots
\ee
where $g^{(0)}_{ab}$ is the zeroth order metric given in
eq.(\ref{zmetric}) and $g^{(1)}_{ab}, g^{(2)}_{ab} ...$ are the first
order, second order etc corrections.

The dilaton satisfies the equation,
\be
\label{dileq}
\nabla^2 \Phi=0.
\ee

Expanding this we find that to order $\epsilon^2$,
\be
\label{exdilb}
\nabla_0^2 \Phi_0 + \nabla_0^2 \Phi_1 + \nabla_1^2 \Phi_0+ \nabla_1^2 \Phi_1
 + \nabla_0^2 \Phi_2=0.
\ee
Here $\nabla_0^2$ is the Laplacian which arises from the zeroth order
metric, and $\nabla_1^2,\nabla_2^2$ are the corrections to the
Laplacian to order $\epsilon, \epsilon^2$ respectively, which arise
due to the corrections in the metric.  The first term on the left
hand side is of order $\epsilon^2$, since it involves two time
derivatives acting on $\Phi_0$. The second term is of order
\footnote{It is easy to see that $\Phi_1$, if non-vanishing, must
depend on the radial coordinate, this makes $\nabla_0^2\Phi_1$ of
order $\epsilon$. $\Phi_1$ would be $r$ dependent for the same reason
that $\Phi_2$ in eq(\ref{dil2}) is.}  $\epsilon$, and so is the third term.
However, we see in \S2.3 that the $O(\epsilon)$ correction to the metric and thus $\nabla_1^2$ vanishes. So the second term is the only one of $O(\epsilon)$ and we learn that
\be
\label{condpa}
\Phi_1=0.
\ee
 The first correction to the dilaton therefore arises at $O(\epsilon^2)$.
Eq.(\ref{exdilb}) now becomes,
\be
\label{2ndorderdil}
\nabla_0^2 \Phi_0 + \nabla_0^2 \Phi_2=0. \ee
Since $\Phi_0$ preserves
the $S^3$ symmetry of $AdS_5$, $\Phi_2$ will also be $S^3$ symmetric
and must therefore only be a function of $t,r$. Further since $\Phi_2$
is  $O(\epsilon^2)$ any time derivative on it would be of higher
order and can be dropped.  Solving eq.(\ref{2ndorderdil}) then gives,
\be
\label{dil2}
 \Phi_2(r,t)= \int^r {dr^\prime \over (r^\prime)^3(1+(r^\prime)^2)}
\left[ \int^{r^\prime} {y^3\over 1+y^2}
dy \ \ \ddot{\Phi}_0(t)+a_1(t) \right]+ a_2(t).
\ee
Here $a_1(t),a_2(t)$ are two
functions of time which arise as integration ``constants".

The integrations in (\ref{dil2}) can be performed, leading to
\bea
\Phi_2 (r,t) &=& \frac{1}{4}  \ddot{\Phi}_0(t) \left[ \frac{1}{r^2}
 \log (1+r^2)
   - \frac{1}{2} (\log(1+r^2))^2 - {\rm dilog} (1+r^2)
  \right]
\nn \\
& & + a_1(t) \frac{1}{2} \left[ \log (1+r^2) - \frac{1}{r^2} - 2 \log~r
  \right] ~+a_2(t).
\eea
The first term in $\Phi_2$ is regular at $r=0$, while the term
multiplying $a_1(t)$ diverges here.
To
find a self-consistent solution in perturbation theory $\Phi_2$ must
be small compared to $\Phi_0$ for all values of $r$, we therefore set
$a_1=0$.
The first term in $\Phi_2 (r,t)$ has the following expansion for large
values of $r$,
\ben
 \ddot{\Phi}_0(t) \left[\frac{\pi^2}{24} - \frac{1}{4r^2} + \left(
   \frac{3}{16} + \frac{1}{4} \log~r \right)\frac{1}{r^4} + \cdots
   \right].
\een
Since we are solving for the dilaton with a specified boundary value
$\Phi_0 (t)$, $\Phi_2(r,t)$ should vanish at the boundary. This
determines $a_2(t)$ to be,
\ben
a_2(t) = -\frac{\pi^2}{24} \ddot{\Phi}_0(t),
\een
leading to the final solution
\ben
\label{dil2f}
 \Phi_2(r,t) =  \frac{1}{4} \ddot{\Phi}_0(t) \left[ \frac{1}{r^2}
\log (1+r^2)
- \frac{1}{2} (\log(1+r^2))^2 - {\rm dilog} (1+r^2) -
  \frac{\pi^2}{6}
  \right].
\een
The solution is regular everywhere. Since ${\rm
  Lim}_{t \rightarrow -\infty} \dot{\Phi}_0 (t), \ddot{\Phi}_0 (t) =
  0$, the correction vanishes in the far past, as required.

\subsection{Corrections to the Metric}

The time varying dilaton provides an additional source of stress energy.
The lowest order contribution due to this stress energy is $O(\epsilon)^2$
as we will see below.  It then follows, after a
suitable coordinate transformation if necessary, that the
$O(\epsilon)$ corrections to the metric vanish and the first
non-vanishing corrections to it arise at order $\epsilon^2$. The
essential point here is that any $O(\epsilon)$ correction to the
metric must be $r$ dependent and thus would lead to a contribution to
the Einstein tensor of order $\epsilon$, which is not allowed. This is
illustrated by the dilaton calculation above, where a similar argument
lead to the $O(\epsilon)$ contribution, $\Phi_1$, vanishing.
In this subsection we calculate  the leading $O(\epsilon^2)$ corrections to
the metric.

Before we proceed it is worth discussing the boundary conditions which
must be imposed on the metric. As was discussed in the previous
subsection we consider a dilaton source, $\Phi_0$, which approaches a
constant value in the far past, $t\rightarrow -\infty$. The
corrections to the metric that arise from such a source should also
vanish in the far past. Thus we see that as $t \rightarrow -\infty$
the metric should approach that of $AdS_5$ space-time.  Also the
solutions we are interested in correspond to the gauge theory living
on a time independent $S^3\times R$ space-time in the presence of a
time dependent Yang Mills coupling (dilaton). This means the leading
behaviour of the metric for large $r$ should be that of $AdS_5$ space.
Changing this behaviour corresponds to turning on a non-normalisable
component of the metric and is dual to changing the metric of the
space-time on which the gauge theory lives.

We expect that these boundary conditions, which specify both the
 behaviour as $t\rightarrow -\infty$ and as $r\rightarrow \infty$
 should lead to a unique solution to the super gravity equations.  The
 former determine the normalisable modes and the latter the
 non-normalisable modes.  This is dual to the fact that in the gauge
 theory the response should be uniquely determined once the time
 dependent Lagrangian is known (this corresponds to the fixing the
 non-normalisable modes) and the state of the system is known in the
 far past(this corresponds to fixing the normalisable modes).

Since $\Phi_0$ is  $S^3$ symmetric, we can consistently assume that the
corrections to the metric will also preserve the  $S^3$ symmetry.
The resulting metric can then be written as,
\be
\label{corrmet}
ds^2=-g_{tt} (t,r) dt^2+g_{rr} (t,r) dr^2+2g_{tr} (t,r)
dtdr+R^2 d\Omega^2.
\ee
Now as is discussed in Appendix A upto $O(\epsilon^2)$ we can consistently
set $g_{tr}=0$. In addition we can to this order set $R^2=r^2$.
Below we also use the notation,
\be
\label{defA}
g_{tt}\equiv e^{2A (t,r),}
\ee
\be
\label{defB}
g_{rr}\equiv e^{2B(t,r)}.
\ee
The metric then takes the form,
\be
\label{fmet}
ds^2=-e^{2A(t,r)} dt^2 + e^{2B(t,r)} dr^2 + r^2 d\Omega^2.
\ee

The trace reversed Einstein equation are:
\be
\label{trrev}
R_{AB}=\Lambda g_{AB} + {1\over 2} \partial_A \Phi \partial_B \Phi.
\ee
In our conventions,
\be
\label{vallambda}
\Lambda=-4.
\ee
To order $\epsilon^2$ we can set $\Phi=\Phi_0$ in the second term on the rhs.

A few simple observations make the task of computing the curvature
components to $O(\epsilon^2)$ much simpler. As we mentioned above the
first corrections to the metric should arise at $O(\epsilon^2)$. To
order $\epsilon^2$ the metric is then
\be
\label{defm}
g_{ab}(t,r) = g_{ab}^{(0)} (r) + g_{ab}^{(2)} (t,r).
\ee
Now the zeroth order metric, $g_{ab}^{(0)}$, is time independent. The
 time derivatives of $g_{ab}^{(2)}$ are non-vanishing but of order
 $\epsilon^3$ and thus can be neglected for calculating the curvature
 tensor to this order. As a result for calculating the curvature
 components to order $\epsilon^2$ we can neglect all time derivatives
 of the metric, eq.(\ref{defm}).

 Before proceeding we note that the comments above imply that the
 equations determining the second order metric components
 schematically take the form,
\be
 \label{sce}
 \hat{O}(r) g_{ab}^{(2)}=f_{ab}(r) \dot{\Phi}_0^2
\ee
where
 $\hat{O}(r)$ is a second order differential operator in the radial
 variable, $r$. As a result the solution will be of the form, \be
 \label{scf}
 g_{ab}^{(2)}={\cal F}(r)_{ab} {\dot\Phi_0}^2,
\ee
where ${\cal F}(r)$
 are functions of $r$ which arise by inverting $\hat{O}(r)$.  We see
 that the corrections to the metric at time $t$ are determined by the
 dilaton source $\Phi_0$ at the same instant of time time $t$.  Note
 also that since we are only considering a dilaton source $\Phi_0$
 which vanishes in the far past, the solution eq.(\ref{scf}) correctly
 imposes the boundary condition that $g_{ab}^{(2)}$ vanishes in far
 past and the metric becomes that of $AdS_5$.

Bearing in mind the discussion above, the curvature components are now
 easy to calculate.  The $t-t$ component of eq.(\ref{trrev}) gives,
 \be
\label{tt}
{(A'e^{(A-B)})'\over e^{(A+B)}}+ 3 {A'e^{-2B}\over
r}={\dot{\Phi}_0^2\over 2}e^{-2A} +4.
\ee
The $r-r$ component gives,
\be
\label{rr}
 -{(A'e^{(A-B)})'\over e^{(A+B)}}+3 {B'e^{-2B}\over r}=-4.
 \ee
 The component with legs along the $S^3$ gives,
 \be
 \label{thetatheta}
 {B'-A'\over e^{2B} r}+{2\over r^2}(1-e^{-2B})=-4.
 \ee
 In these equations primes indicates derivative with respect to $r$
 and dot indicates derivative with respect to time.

 Adding the $t-t$ and $r-r$ equations gives,
 \be
 \label{addtr}
 3(A'+B'){e^{-2B}\over r}={\dot{\Phi}_0^2\over 2}e^{-2A}.
\ee
 Eq.(\ref{thetatheta}) and eq.(\ref{addtr}) then lead to

\be
 \label{rese}
 {2B'e^{-2B}\over r}-{1\over 6}\dot{\Phi}_0^2e^{-2A}+{2\over
 r^2}(1-e^{-2B})=-4.  \ee This is a first order equation in
 $B$. Integrating we get to order $\epsilon^2$, \be
 \label{intb}
 e^{-2B}=1+r^2+{c_1\over r^2}-{1\over 6}{\dot{\Phi}_0^2\over
 r^2}[\int_0^r e^{-2A_0}r^3 dr].
\ee
Here $c_1$ is an integration
 constant and $e^{2A_0}=1+r^2$ is the zeroth order value of
 $e^{2A}$. We require that the metric become that of $AdS_5$ space as
 $t\rightarrow -\infty$ this sets $c_1=0$ \footnote{Note that $c_1$
 could be a function of time and still solve eq.(\ref{rese}), recall
 though that the equations above were derived by neglecting all time
 derivatives of the metric, eq.(\ref{defm}). Only a time independent
 constant $c_1$ is consistent with this assumption. A similar argument
 will also apply to the other integration constants we obtain as we
 proceed.}. A negative value of $c_1$ would mean starting with a black
 hole in $AdS_5$ in the far past.

The integral within the square brackets on the rhs in eq.(\ref{intb})
 is given by,
\be
 \label{integral}
 \int_0^r e^{-2A_0}r^3dr={1\over 2}[r^2- \ln(1+r^2) + d_1].
 \ee
This gives,
\be
\label{fB}
e^{-2B}=1+r^2-{1\over 12}{\dot{\Phi}_0^2\over r^2}[r^2- \ln(1+r^2) + d_1].
\ee
 A solution which is regular for all values of $r$,
  is obtained by setting $d_1$ to vanish.
 This gives,
 \be
 \label{feB}
 e^{-2B}=1+r^2-{1\over 12}\dot{\Phi}_0^2[1-{1\over r^2}\ln(1+r^2)].
 \ee

 We can obtain $e^{2A}$ from eq.(\ref{addtr}).  To second order in
 $\epsilon^2$ this equation becomes, \be
 \label{a2order}
 A'={1\over 6}r \dot{\Phi}_0^2e^{-2(A_0-B_0)}-B',
 \ee
which gives,
\be
\label{solA}
A=-B+{1\over 12}\dot{\Phi}_0^2[-{1\over 1+r^2}+d_3],
\ee
with $d_3$ being a general function of time.
Eq.(\ref{solA}) and eq.(\ref{feB}) leads to \be
 \label{nfeA}
 e^{2A}=1+r^2 +\dot{\Phi}_0^2[-{1\over 4} + {1\over 12}
 {\ln(1+r^2)\over r^2}+{d_3\over 6}(1+r^2)].
\ee
The last term on
 the right hand side changes the leading behaviour of $e^{2A}$ as
 $r\rightarrow \infty$, if $d_3$ does not vanish, and therefore
 corresponds to turning on a non-normalisable mode of the metric. As
 was discussed above we want solutions where this mode is not turned
 on, and we therefore set $d_3$ to vanish.

 This gives finally, \be
 \label{feA}
 e^{2A}=1+r^2-{1\over 4} \dot{\Phi}_0^2+{1\over 12}\dot{\Phi}_0^2
 {\ln(1+r^2)\over r^2}.
\ee
Eq.(\ref{feB}), (\ref{feA}) are the
 solutions to the metric, eq.(\ref{fmet}), to second order.  Note that
 the Einstein equations gives rise to three equations, eq.(\ref{tt}),
 eq.(\ref{rr}), eq.(\ref{thetatheta}).  We have used only two linear
 combinations out of of these to find $A,B$. One can show that the
 remaining equation is also solved by the solution given above.

 In summary we note that the Einstein equations can be solved
 consistently to second order in $\epsilon^2$. The resulting solution
 is horizon-free and
regular for all values of the radial coordinate and satisfies the
 required boundary conditions discussed above.  The second order
 correction to the metric is parametrically suppressed by $\epsilon^2$
 compared to the leading term for all values of $r$, thereby making
 the perturbation theory self consistent.

 Let us end by commenting on the choice of integration constants made
 in obtaining the solution above. The boundary conditions, as
 $t\rightarrow -\infty$ and $r\rightarrow \infty$, determine most of
 the integration constants. One integration constant $d_1$ which
 appears in the solution for $e^{2B}$, eq.(\ref{fB}) is fixed by
 regularity at $r\rightarrow 0$ \footnote{Similarly in solving for the
 dilaton perturbation the integration constant $a_1$ is fixed by
 requiring regularity at $r=0$, eq.(\ref{dil2}).}.  For $d_1=0$ the
 second order correction is small compared to the leading term, and
 the use of perturbation theory is self-consistent.  Moreover we
 expect that the boundary conditions imposed here lead to a unique
 solution to the supergravity equations, as was discussed at the beginning of
 this subsection. Thus the solution obtained by setting $d_1=0$ should
 be the correct one.

The solution above is regular and has no horizon. It has these properties due to the slowly varying nature of the boundary dilaton. The dual field theory in this case is in a non-dissipative phase. Once the dilaton begins to change sufficiently rapidly with time we expect that a black hole is formed, corresponding to the formation of a strongly dissipative phase in the dual field theory.
 In \cite{Bhattacharyya:2009uu} the
effect of a {\em small amplitude} time dependent dilaton with
arbitrary time dependence was studied. Indeed it was found that when the time variation is fast
enough there are no regular horizon-free solutions and a black hole is
formed.

Finally, the analysis of this section holds when $e^\Phi$ is large
enough to ensure applicability of supergravity. The fact that a black
hole is not formed in this regime does not preclude formation of black
holes from stringy effects when $e^\Phi$ becomes small enough. In fact we will argue in later sections
that the latter is a distinct possibility.

\subsection{Effective decoupling of modes}

An important feature of the lowest order calculation of this
section is that the perturbations of the dilaton and the metric are
essentially linear and do not
couple to each other. To this order, the dilaton perturbation is
simply a solution of the linear d'Alembertian equation in
$AdS_5$. Similarly the metric perturbations also satisfy the
linearized equations of motion in $AdS$,
albeit in the presence of a source
provided by the energy momentum tensor of the dilaton. This is a
feature present only in the leading order calculation. As explained
above, this arises because of the smallness of the parameter
$\epsilon$. We will use this feature to compare leading order
supergravity results with gauge theory calculations in a later section.

\section{Calculation of Stress Tensor and Other Operators}

In this section we calculate the boundary
stress tensor and the expectation value of the operator dual to the
dilaton, staying in the supergravity approximation. This will be done
using standard techniques of holographic renormalization group
\cite{skenderis,balasubramanian, Nojiri:1998dh,
Emparan:1999pm,Awad:1999xx, Haro, Nojiri2, Fukuma:2002sb}.

\subsection{The Energy-Momentum Tensor}

The metric is of the form, eq.(\ref{fmet}), eq.(\ref{feB}),
eq.(\ref{feA}). For calculating the stress tensor a boundary is
introduced at large and finite radial location, $r=r_0$.  The induced
metric on the boundary is,
\be
\label{indmet}
ds_B^2\equiv h_{\mu\nu}dx^\mu
dx^\nu=-e^{2A}dt^2+r^2(d\theta^2+\sin^2\theta d\phi^2+\sin^2\theta
\sin^2\phi d\psi^2).
\ee
The 5 dim. action is given by
\be
\label{5dact}
S_5={1\over 16\pi G_5}\int_Md^5x\sqrt{-g}\bigl(R+12-{1\over 2}(\nabla
\Phi)^2\bigr) -{1\over 8 \pi G_5}\int_{r=r_0} d^4x\sqrt{-h} \Theta.
\ee
Here $h_{\mu\nu}$ is the induced metric on the boundary , and
$\Theta$ is the trace of the extrinsic curvature of the boundary.  In our
conventions, with $R_{AdS}=1$,
\be
\label{valG5}
G_5={\pi\over 2 N^2}.
\ee
A counter term needs to be added, it is,
\be
\label{cterm}
S_{ct}=-{1 \over 8 \pi G_5}\int_{\partial M}d^4x\sqrt{-h}\bigl
[3+{{\cal R}\over 4} -{1\over 8}(\nabla \Phi)^2-log(r_0) a_{(4)} \bigr
].
\ee

The last term is needed to cancel logarithmic divergences which arise
in the action, it is well known and is discussed in e.g.
\cite{skenderis, Nojiri2}. From eq.(24) of \cite{Nojiri2} we have
\footnote{Note that our definition of the dilaton $\Phi$ is related to
$\phi_{(0)}$ in \cite{Nojiri2} by $\phi_{(0)}=\Phi/2$.} that
\ben
\label{vala4}
a_{(4)} =  \ {1\over 8}
R_{\mu\nu}R^{\mu\nu}-{1\over 24} R^2 -{1\over
8}R^{\mu\nu} \partial_\mu\Phi \partial_\nu\Phi+{1\over 24}R
h^{\mu\nu}\partial_\mu \Phi \partial_\nu \Phi +{1\over
16} (\nabla^2\Phi)^2+{1\over 48}\{(\nabla \Phi)^2\}^2 \ \ .
\een
Here $\nabla$ is a covariant derivative with respect to the metric
$h_{\mu\nu}$.

 Varying the total action $S_T=S_5+S_{ct}$ gives the stress
energy,
\bea
\label{bstressa}
T^{\mu\nu}& = & {2\over \sqrt{-h}}{\delta S_T\over \delta
          h_{\mu\nu}}\\ \nonumber
           &=& {1\over 8 \pi
          G_5}\bigl[\Theta^{\mu\nu}-\Theta h^{\mu\nu}-3h^{\mu\nu}  +{1\over 2}
          G^{\mu\nu} -{1\over 4}\nabla^\mu \Phi \nabla^\nu \Phi+{1\over 8}
          h^{\mu\nu} (\nabla \Phi)^2 + \cdots \bigr].
\end{eqnarray}
Here $G^{\mu\nu}$ is the Einstein tensor with respect to the metric $h_{\mu\nu}$.
The ellipses stand for extra terms obtained by varying the last term in eq.(\ref{cterm}) proportional to $a_{(4)}$. While these terms are not explicitly written down in   eq.(\ref{bstressa}), we do include them  in the calculations below.

The expectation value of the stress tensor in the boundary theory is
then given by, \be
\label{exstress}
<T^\mu_\nu>=r_0^4 T^\mu_\nu
\ee
Carrying out the calculation gives a finite answer,
\begin{eqnarray}
\label{finalst}
<T^t_t>&=&{N^2\over 4\pi^2}[-{3\over 8}-{\dot{\Phi}_0^2\over 16}]\cr
<T^\theta_\theta>=<T^\psi_\psi>=<T^\phi_\phi>&=&{N^2\over
4\pi^2}[{1\over 8}-{\dot{\Phi}_0^2\over 16}]
\end{eqnarray}
where we have used eq.(\ref{valG5}).  We remind the reader that in our
conventions the radius of the $S^3$ on which the boundary gauge theory
lives has been set equal to unity.  The first term on the right hand
side of (\ref{finalst})
arises due to the Casimir effect. The second term is the additional
contribution due to the varying Yang Mills coupling.

From eq.(\ref{finalst}) the total energy in the boundary theory can be
calculated. We get,
\be
\label{fmass}
E=-<T^t_t> V_{S^3}={3N^2\over 16}+{N^2\dot{\Phi}_0^2\over 32}.  \ee
where $V_{S^3}=2\pi^2$ is the volume of a unit three-sphere.  Note
that the varying dilaton gives rise to a positive contribution to the
mass, as one would expect. Moreover this additional contribution
vanishes when the $\dot{\Phi}$ vanishes. In particular for a dilaton
profile which in the far future, as $t\rightarrow \infty$, again
approaches a constant value (which could be different from the
starting value it had at $t \rightarrow -\infty$) the net energy
produced due to the varying dilaton vanishes.

 \subsection{Expectation value of the Operator Dual to the Dilaton}

 The operator dual to the dilaton has been discussed explicitly in \cite{Witten1},
 \cite{ken1}, \cite{adnt}.

It's expectation value is given by, \be
 \label{expdil}
 <\hat{{\cal O}}_{l=0}>={\delta S_T\over \delta \Phi_B}|_{\Phi_B\rightarrow 0}
 \ee
 Here $S_T$ is the total action including the boundary terms, eqn. (\ref{cterm}).
Since $\Phi_B$ is a function of $t$ alone the lhs is the $l=0$ component of the operator
dual to the dilaton which we denote by, $\hat{{O}}_{l=0}$.

The steps involved are analogous to those above for the stress tensor
  and yield,
\be
  \label{exdilf}
  <\hat{{\cal O}}_{l=0}>=-{N^2\over 16}\ddot{\Phi}_0
  \ee
Note that the lhs refers to the expectation value for the dual
  operator integrated over the boundary $S^3$.  In obtaining
  eq.(\ref{exdilf}) we have removed all the divergent terms and only
  kept the finite piece. A quadratically divergent piece is removed by the third term in eq.(\ref{cterm}) proportional to $(\nabla \Phi)^2$, and a log divergence is removed by a contribution from the last term in eq.(\ref{cterm}) proportional to $a_{(4)}$.

\subsection{Additional Comments}
Let us end this section with a few comments.

   The only source for time dependence in the boundary theory is the
   varying Yang Mills coupling.  A simple extension of the usual
   Noether procedure for the energy, now in the presence of this time
   dependence, tells us that
\be
   \label{noether}
   {dE\over dt}=-\dot{\Phi}_0 <\hat{{\cal O}}_{l=0}>.  \ee
It is easy to see that the
   answers obtained above in eq.(\ref{fmass}), eq.(\ref{exdilf})
   satisfy this relation.  The relation eq.(\ref{noether}) is a
   special case of a more general relation which applies for a dilaton
   varying both in space and time, this was discussed in Appendix A of
\cite{BLMNTW}.

In general, for a slowly varying dilaton one can expand $<\hat{{\cal O}}_{l=0}>$ in
a power series in $\dot{\Phi}_0$. For constant dilaton, the solution is
$AdS_5$ where one knows that the $<\hat{{\cal O}}_{l=0}>$ vanishes. Thus one can
write,
\be
\label{expandd}
<\hat{{\cal O}}_{l=0}>=c_1 \dot{\Phi}_0+c_2\ddot{\Phi}_0 + c_3 (\dot{\Phi}_0)^2 \cdots
\ee
where the ellipses stand for higher powers of derivatives of the
dilaton.  Comparing with the answer in eq.(\ref{exdilf}) one sees that
in the supergravity limit $c_1$ and $c_3$ vanish. As a result
${dE\over dt}$ is a total derivative, and as was discussed above if
the dilaton asymptotes to a constant in the far future there is no net
gain in energy.

It is useful to contrast this with what happens in the case of an
infinite black brane at temperature $T$ subjected to a time dependent
dilaton which is slowly varying compared to the temperature $T$. This
situation was analysed extensively in \cite{BLMNTW}.
In that case (see
eq.(2.13), eq.(3.20) and section 7.2 of the paper) the leading term in
eq.(\ref{expandd}) proportional to $\dot{\Phi}_0$ does not vanish.  The
temperature then satisfies an equation, \be
\label{tempeq}
{dT\over dt}={1\over 12 \pi} \dot{\Phi}_0^2
\ee
As a result any variation in the dilaton leads to a net increase in
the temperature, and the energy density.
Note the first term in eq.(\ref{expandd}) contains only one derivative with respect to time and breaks time reversal invariance. It can only arise in a dissipative system. In the case of a black hole  the formation of a horizon breaks time reversal invariance and turns the system dissipative allowing   this term to arise. In the solution we construct no horizon forms and consistent with that the first term is absent.

We see in the solution discussed above that the second order corrections to the dilaton and metric arise in an instantaneous manner - at some time $t$, and for all values of $r$, they are determined by the boundary value of the dilaton at the same instant of time $t$. This might seem a little puzzling at first since  one would have expected the effects of the changing boundary conditions to be felt in a retarded manner. Note though that in  AdS space a light ray can reach any point in the bulk from the boundary within  a time of order $R_{AdS}$.  When $\epsilon \ll 1$ this is much smaller than the time taken for the boundary conditions to change appreciably. This  explains why the leading corrections arise in an instantaneous manner. Some of the corrections which arise at higher order would turn this instantaneous response into a  retarded one.

From the solution and the expectation values of the energy and
$\hat{{\cal O}}_{l=0}$ it follows that in the far future the system settles down into an $AdS_5$ solution again. The near instantaneous nature of the solution means that this happens quickly on the times scale of order $R_{AdS}$. This agrees with general expectations. The supergravity modes carry an energy of order $1/R_{AdS}$ and should give rise to a response time of order $R_{AdS}$.

Also note that in our units, where $R_{AdS}=1$, each supergravity mode carries an energy of order unity.  The total energy at intermediate times is of order $N^2\epsilon^2$, so we see that an $O(N^2\epsilon^2)$ number of quanta are excited by the time varying boundary dilaton. This can be a big number when $N\epsilon \gg 1$. In fact the energy is really carried by the various dilaton modes. The metric perturbations are $S^3$ symmetric and thus contain no gravitons (in the sense of genuine propagating modes). One can think of this energy as being  stored  in a spatial region of order $R_{AdS}$ in size located at the center of AdS space. This is what one would expect, since the supergravity modes which are produced by the time varying boundary dilaton have a size of order $R_{AdS}$ and their gravitational redshift  is biggest at the center of AdS space \footnote{AdS is of course a homogeneous space-time, but our boundary conditions pick out a particular notion of time. The center of AdS, where the energy is concentrated, is the region  as mentioned above where the    redshift  in the corresponding  energy is the biggest.}.

In summary, the response in the bulk to the time varying boundary dilaton is characteristic of a non-dissipative  adiabatic system which is being driven much more slowly than  its own fast  internal time scale of  response.

\section{Gauge Theory : Quantum  Adiabatic Approximation}

We now turn to analysing the behaviour of the system in the dual field
theory.  The motivation behind this is to be able to extend our
understanding to situations in which the 'tHooft coupling at
intermediate time becomes of order one or smaller, so that the
geometry in the bulk becomes of order the string scale. In such
situations the supergravity calculation presented in the previous
section breaks down and higher derivative corrections become
important. The gauge theory description continues to be valid,
however. Using this description one can then hope to answer how the
system evolves in the region of string scale curvature, and in
particular whether by waiting for enough time a smooth geometry with
small curvature emerges again on the gravity side.

We saw in the previous subsection that the bulk response was
characteristic of an adiabatic system which was being driven slowly
compared to the time scale of its own internal response. This suggests
that in the gauge theory also an adiabatic perturbation theory should
be valid and should prove useful in understanding the response. A
related observation is the following.  The bulk solutions we have
considered correspond to keeping the radius $R$ of the $S^3$ on which
the gauge theory lives to be constant and independent of time.  We
will choose conventions in which $R=R_{AdS}=1$.  The Yang Mills theory
is related to the boundary dilaton by, \be
\label{gtdil}
g_{YM}^2=e^{\Phi_0(t)}.
\ee
The dilaton profile eq.(\ref{redil}) also means that Yang Mills
coupling in the gauge theory varies slowly compared to the radius
$R$. Since this is the only other scale in the system, this also
suggests that an adiabatic approximation should be valid in the
boundary theory.

We will discuss two different kinds of adiabatic perturbation theory
below.  The first, which we call Quantum adiabatic perturbation
theory, is studied in this section. This is the adiabatic perturbation
theory one finds discussed in a standard text book of quantum
mechanics, see \cite{Migdal1},\cite{Migdal:1977bq}.  Its validity, we
will see below requires the condition, $N \epsilon \ll 1$, to be
met. We will argue that once this condition is met the gauge theory
analysis allows us to conclude that, even in situations where the
curvature becomes of order the string scale at intermediate times, a
dual smooth $AdS_5$ geometry emerges as a good approximation in the
far future.

The supergravity calculations, however, required only
the condition $\epsilon \ll 1$, which
is much less restrictive than the condition $N
\epsilon \ll 1$. Understanding the supergravity regime on the gauge theory
side leads us to formulate another perturbation theory, which we call
``Large N Classical Adiabatic Perturbation Theory'' (LNCAPT).
To explain this we find
it useful to first discuss the example of a driven harmonic
oscillator, as considered in \S5. Following this, we discuss
LNCAPT in the gauge theory in \S6. We
find that its validity requires that the conditions
eq.(\ref{condcap1}) are met.  Using it we will get agreement with the
supergravity calculations of sections \S2, \S3, when the 'tHooft
coupling remains large for all times.

Towards the end of \S6, we discuss what happens in the gauge theory
when conditions eq.(\ref{condcap1}) are met but with the 'tHooft
coupling becoming small at intermediate times. Two qualitatively
different behaviours are possible, and we will not be able to decide
between them here.  Either way, at late times a mostly smooth AdS
description becomes good on the gravity side, with the possible
presence of a small black hole.

In the discussion below we will consider the following type of profile  for the boundary dilaton: it  asymptotes in the far past and future to constant values such that the initial and final values of the 'tHooft coupling, $\lambda$,  are big, and attains its minimum value near $t=0$.
If this minimum value of $\lambda \le 1$ the supergravity approximation will break down.
We will also take the initial state of the system to be the ground state of the ${\cal N}=4$ theory, on $S^3$ the spectrum of the gauge theory is gapped and this state is well defined.

\subsection{The Quantum Adiabatic Approximation}
\subsubsection{General Features}

It is well known that the spectrum of the ${\cal N}=4$ theory on $S^3$ has a gap between the energy of the lowest state and the first excited state.
This gap is of order $1/R$ and thus  is of order unity in our conventions.
The existence of this gap follows very generally just from the fact that
the spectrum must provide a unitary representation of the conformal group, \cite{Mack},
and the gap is therefore present   for {\em all } values of the  Yang Mills
coupling constant. In the supergravity approximation the spectrum can be
calculated using the gravity description and is consistent with the gap,
the lowest lying states have an energy $E=2$. This is also true at
very weak 'tHooft coupling.

Now for a slowly varying dilaton eq.(\ref{redil}) we see that the Yang Mills
coupling and therefore the externally imposed time dependence varies slowly compared
to this gap. There is a well known  adiabatic approximation which is known to work in
such situations, see e.g.~\cite{Migdal1},\cite{Migdal:1977bq} whose treatment
we closely follow. We will refer to this as the quantum  adiabatic approximation below and study the Yang Mills theory in this approximation.

The essential idea behind this approximation is that when a system is subjected to a time dependence which is slow  compared to its internal response time, the system can adjust itself very quickly and as a result to good approximation stays in the ground state of the instantaneous Hamiltonian.

More precisely,
consider a time dependent Hamiltonian
$H(\llambda(t))$, where $\llambda(t)$ is the time varying
parameter. Now consider the one parameter family of time {\it
independent} Hamiltonians given by $H(\llambda)$. To make our notation
clear, a different value of $\llambda$ corresponds to a different
Hamiltonian in this family, but each Hamiltonian is time independent.
Let $|\phi_m(\llambda)>$ be a complete set of eigenstates of the
Hamiltonian $H(\llambda)$ satisfying,
\be
\label{evlh}
H(\llambda) |\phi_m(\llambda)>=E_m(\llambda) |\phi_m(\llambda)>,
\ee
in particular let the ground state of $H(\llambda)$ be given by
$|\phi_0(\llambda)>$. We take $|\phi_m(\llambda)>$ to have unit
norm. Then the adiabatic theorem states that if $\llambda
\rightarrow \llambda_0$ in the far past, and we start with the state
$|\phi_0>$ which is the ground state of $H(\llambda_0)$ in the far
past, the state at any time $t$ is well approximated by,
\be
 \label{zaps}
 |\psi^0(t)>\simeq |\phi_0(\llambda)> e^{-i\int_{-\infty}^t
 E_0(\llambda) dt}.  \ee
Here $|\phi_0(\llambda)>$ is the ground state
 of the time independent Hamiltonian corresponding to the value
 $\llambda=\llambda(t)$. Similarly in the phase factor $E_0(\llambda)$
 is the value of the ground state energy for $\llambda=\llambda(t)$.

Corrections can be calculated by
 expanding the state at time $t$ in a basis of energy eigenstates at
 the instantaneous value of the parameter $\llambda$.  The first
 corrections take the form,
\be
 \label{fcorr}
 |\psi^1(t)>=\sum_{n\ne 0} a_n(t)
 |\phi_n(\llambda)>e^{-i\int_{-\infty}^t E_n dt} \ee where the
 coefficient $a_n(t)$ is, \be
\label{valan}
a_n(t)=-\int_{-\infty}^t dt'{<\phi_n(\llambda)|{\partial H\over
\partial \llambda}|\phi_0(\llambda)>\over E_0-E_n} \ \dot{\llambda} \
e^{-i\int_{-\infty}^{t'}(E_0-E_n)dt'} \ee
In the formula above on the
rhs $|\phi_n(\llambda)>, {\partial H\over \partial \llambda},
E_n(\llambda),$ are all functions of time, through the time dependence
of $\llambda$.

\subsubsection{Conditions For Validity}

For the adiabatic approximation to be good the first corrections must
be small. To ensure this  we impose the condition,
\be
\label{condada}
|<\phi_n|{\partial H \over \partial \zeta}|\phi_0> \dot{\llambda}| \ll
 (E_1-E_0)^2
\ee
where $(E_1-E_0)$ is the energy gap between the ground state and the
first excited state and $|\phi_n>$ is any excited state.  (This would
 then imply that the lhs in eq.(\ref{condada}) is smaller than $(E_n-E_0)^2$ for
 all $n$.)
This
condition is imposed  for all time for the adiabatic approximation to be
valid \footnote{The actual condition is that the corrections to
$|\psi^0>$ must be small. This means that at first order
$<\psi^1|\psi^1>$ should be small. When eq.(\ref{condada}) is met
$|a_n|$ is small, but in some cases that might not be enough and the
requirement that the sum $\sum |a_n|^2$ is small imposes extra
restrictions. There could also be additional conditions which arise at
second order etc.} .

In our case the role of the parameter $\llambda$ is played by the
dilaton $\Phi_0$(with the gauge coupling $g_{YM}^2=e^{\Phi_0}$). Thus
eq.(\ref{condada}) takes the form,
\be
\label{dercondth}
|<\phi_n|{\partial H \over \partial \Phi_0}|\phi_0> \dot{\Phi_0}|
\ll (E_1-E_0)^2.
\ee
Now, as we will see below in subsection \S4.3, ${\partial H \over
\partial \Phi_0}$ is, up to a sign, exactly the operator $\hat{{\cal
O}}_{l=0}$ which is dual to the modes of the dilaton
which are spherically symmetric on the $S^3$.
Therefore eq.(\ref{dercondth}) becomes
\be
\label{condresta}
|<\phi_n|\hat{{\cal O}}_{l=0}|\phi_0> \dot{\Phi_0}| \ll (E_1-E_0)^2.
\ee

We have argued above that the rhs is of order unity in our conventions
 due to the existence of a robust gap. On the lhs, $\dot{\Phi_0} \sim
 O(\epsilon)$, and as we will argue below the matrix element,
 $|<\phi_n|\hat{{\cal O}}_{l=0}|\phi_0> \sim O(N)$.  Thus
 eq.(\ref{condresta}) becomes,
\be
\label{condrz}
N \epsilon \ll 1.
\ee

\subsection{Highly Curved Geometries}

Eq.(\ref{condrz}) is the required condition then for the applicability
of quantum perturbation theory. When this condition is met, we can
continue to trust the quatum adiabatic approximation in the gauge
theory  even when the 'tHooft
coupling becomes of order unity or smaller at intermediate times. All
the conditions which are  required for the validity of this approximation
 continue to be hold in this case. First, as was
discussed above the gap of order unity continues to exist. Second, the
matrix elements which enter are in fact independent of $\lambda$ since
they correspond to the two-point function of dilaton which is a chiral
operator. Thus the system continues to be well described in the
quantum adiabatic approximation so long as eq.(\ref{condrz}) is met.
It follows then that in the far future the state of the system to good
approximation is the ground state of the ${\cal N}=4$ theory. This
implies that the dual description in the far future is a smooth
$AdS_5$ geometry.

There is one important caveat to the above conclusion. It is possible
   that at $\lambda\sim O(1)$ there are several states in the
   spectrum, scaling as a positive power of $N$, which accumulate near
   the first excited state. This does not happen for $\lambda \gg 1$
   and for $\lambda \ll 1$ (where the spectrum of the free theory is
   of course known) but it remains a logical possibility.  If this is
   true the conditions for the adiabatic approximation will have to be
   revised so that the dilaton varies even more slowly as a power of
   $N$.  This is a question which can be settled in principle once the
   spectrum of the ${\cal N}=4$ theory is known for all
   $\lambda$. Similarly, the possibility for unexpected surprises at
   higher orders can also be examined once enough is known about the
   ${\cal N}=4$ theory. The point is simply that in this approximation all
   matrix elements and conditions can be phrased as statements in the
   time independent ${\cal N}=4$ theory. As  our knowledge of the ${\cal N}=4$ theory grows we will be able to check for any  such unexpected surprises.

  Let us also mention before proceeding that when the condition
  eq.(\ref{condrz}) is met and for a dilaton profile where the 'tHooft
  coupling stays large for all time, the metric is to good
  approximation smooth $AdS_5$ for all time. However the small
  corrections to this metric and dilaton cannot be calculated reliably
  in the classical approximation used in section $2$. This is because
  in this regime it is very difficult to even produce one supergravity
  quantum as an excitation above the adiabatic vaccum. Therefore
  quantum effects are important in calculating these corrections.

\subsection{More Comments}
We close this section by discussion two points relevant to the
analysis leading up to condition, eq.(\ref{condrz}).

First, let us argue why ${\partial H \over \partial \Phi_0}
=-\hat{{\cal O}}_{l=0}$. The argument is sketched out below, more
details can be found in \cite{adnt}.  The action of the ${\cal N}=4$
theory is given by,
\be
\label{actn4}
S=\int dt~d\Omega_3~ \sqrt{-g}(-{1\over 4 e^{\Phi_0}})TrF_{\mu\nu}F^{\mu\nu} + \cdots
\ee
where the ellipses indicate extra terms coming from scalars and fermions.
Varying with respect to $\Phi_0$ gives us the operator dual to the dilaton,
\be
\label{dualop}
\hat{{\cal O}}=\sqrt{-g}({1\over 4 e^{\Phi_0}})TrF_{\mu\nu}F^{\mu\nu} +
\cdots
\ee
where the ellipses denote extra terms which arise from the
terms left out in eq.(\ref{actn4}). Henceforth, to emphasise the key
argument we neglect the additional terms coming from the ellipses.

Working in $A_0=0$ gauge, the Hamiltonian  density ${\cal H}$ is given by,
\be
\label{hamiltonian}
{\cal H} = e^{\Phi_0}{\pi_i \pi^i\over 2} + {e^{-\Phi_0} \over 4} F_{ij}F^{ij}
\ee
where
\be
\label{valmom}
\pi_i=e^{-\Phi_0} \partial_0 A_i
\ee
is the momentum conjugate to $A_i$.
Varying with respect to $\Phi_0$  gives,
\be
\label{valopdual}
{\partial {\cal H} \over \partial \Phi_0}
={\pi_i \pi^i \over 2} e^{\Phi_0}-{e^{-\Phi_0}\over 4} F_{ij}F^{ij}.
\ee
Substituting from eq.(\ref{valmom}) one sees that this agrees (up to a
sign) with the operator $\hat{{\cal O}}$ given in eq.(\ref{dualop}).
When the dilaton depends on time alone we can integrate the above
equations over $S^3$, which leads to the relation
${\partial H \over \partial
\Phi_0}=-\hat{{\cal O}}_{l=0}$, where $H$ now stands for
the hamiltonian (rather than the hamiltonian density).

Second, we estimate how the matrix element, $<\phi_n|\hat{{\cal
 O}}_{l=0}|\phi_0>$, which appears in eq.(\ref{condresta}), scales with $N$.
 It is useful to first recall that the ${\cal N}=4$ theory, which is
 conformally invariant, has an operator state correspondence. The
 states $|\phi_n>$ can be thought of as being created from the vacuum
 by the insertion of a local operator. This makes it clear that the
 only states having a non-zero matrix element, $<\phi_n|\hat{{\cal
 O}}_{l=0}|\phi_0>$, are those which can be created from the vacuum by
 inserting $\hat{{\cal O}}_{l=0}$, since the only operator with which
 $\hat{{\cal O}}_{l=0}$ has a non-zero two point function is $\hat{{\cal
 O}}_{l=0}$ itself.

Now in terms of powers of $N$ the two-point function scales like, \be
\label{twopt}
<\hat{{\cal O}}_{l=0} ~~\hat{{\cal O}}_{l=0} >\sim N^2.  \ee The state
$|\phi_n>$ which appears in the matrix element in eq.(\ref{condresta})
has unit norm and is therefore created from the vacuum by the
operator, \be
\label{createphin}
|\phi_n>\sim {1\over N} \hat{{\cal O}}_{l=0} |0> \ee From
eq.(\ref{twopt}), eq.(\ref{createphin}), we then see that the matrix
element scales like, \be
\label{scales}
<\phi_n|\hat{{\cal O}}_{l=0}|\phi_0>\sim N
\ee
as was mentioned above.

Our discussion leading up to the estimate of the matrix element has
been imprecise in some respects.  First, strictly speaking the
operator state correspondence we used is a property of the Euclidean
theory on $R^4$, where as we are interested in the Minkowski theory on
$S^3\times R$. However, this is a technicality which can be taken
care of by first relating the matrix element in the Minkowski theory
to that in Euclidean $S^3\times R$ space and then relating the latter
to that on $R^4$ by a conformal transformation.

More importantly, the state created by $\hat{{\cal O}}_{l=0}$ is not
an eigenstate of energy, but is in fact a sum over an
infinite number of states labelled by an integer $n$ with energies
$\omega_n = 4 + 2n$.
This can be understood as follows. The
operator $\hat{{\cal O}}$ can be expanded into positive and negative
frequency modes, $A_n, A_n^\dagger$ respectively, for an infinite set
$n$, and acting with any of the $A_n^\dagger$'s gives a state, \be
\label{newdefphin}
|\varphi_n> \simeq A_n^\dagger |0>.
\ee
One must therefore worry about the dependence on the mode number $n$
in the matrix element and the effects of summing up the contributions
for all these modes.  We will return to address this issue in more
detail in subsections 6.2 and 6.3, when we describe the operators $A_n,
A_n^\dagger$ more explicitly and discuss renormalization. For now, let
us state that after the more careful treatment we will find that the
condition for the quantum adiabatic approximation eq.(\ref{condrz})
goes through unchanged. The physical reason is simply this: we are
interested here in the very low-frequency response of the system and
its very high frequency modes are not relevant for this.

\section{The Slowly Driven Harmonic Oscillator}

The supergravity calculations required the condition $\epsilon \ll 1$. To
understand this regime in the dual gauge theory it is first useful to
consider a quantum mechanical Harmonic oscillator with
frequency $\omega_0$
driven by a time
dependent source $J(t)$ .
We will see that in this case a
classical adiabatic perturbation theory becomes valid when\footnote{Eq.(\ref{oscconda}),
(\ref{osccondb}), clearly cannot hold when $\dot{J}$ vanishes. The more
precise versions of these
conditions are as follows. Eq.(\ref{oscconda}) is really the requirement that  $J$
is slowly varying.  By this one  means that
the fourier transform of $J$ has support, up to say exponentially small corrections,
 only for small frequencies   compared to $\omega_0$. Eq.(\ref{osccondb}) is the requirement
 that the coherent state parameter, $\lambda(t)$ given in
  eq.(\ref{cstatepara}), is large. }
\be
\label{oscconda}
 {\ddot{J}\over \dot{J} \omega_0}\ll 1,
 \ee
 \be
 \label{osccondb}
 \dot{J}\gg \omega_0^{5/2}.
\ee
Having understood this system we then return
to the gauge theory in the following subsection.

The Hamiltonian is given by
\be
\label{tdef}
H={1\over 2} \dot{X}^2+{1\over 2} \omega_0^2 ( X + {J(t)\over
\omega_0^2})^2.
\ee
In the quantum adiabatic approximation one
considers the instantaneous Hamiltonian. At time $t_0$ this is given
by,
\be
\label{ttwodef}
H_0={1\over 2} \dot{X}^2+{1\over 2} \omega_0^2 (X +
\frac{J(t_0)}{\omega_0^2}
)^2
\ee
where $J(t_0)$ is to regarded as a time independent constant in $H_0$.

The ground state of $H_0$ is a coherent state.  Define, \be
 \label{creatdestruct}
 X={a+a^\dagger\over \sqrt{2\omega_0}}, \ \
 P=-i\sqrt{\omega_0}({a-a^\dagger\over \sqrt{2}}) \ee to be the
 conventional creation and destruction operators. Here, \be
 \label{defmom}
 P=\dot{X}
 \ee
 is the conjugate momentum.
 The ground state is
 \be
 \label{agndstate}
 |\phi_0>=N_\alpha e^{\alpha a^\dagger}|0>. \ee Here $N_\alpha$ is a
 normalisation constant, determined by requiring that $<\phi_0|\phi_0>=1$.
 The state $|0>$ is the vacuum annihilated by $a$, i.e.,
 \be
 \label{vaccumcon}
 a|0>=0,
 \ee
 and
 \be
 \label{valpara}
 \alpha=-{J \over \sqrt{2 \omega_0^3}}.
 \ee
 The ground state energy is
 \be
 \label{gndstateen}
 E_0={1\over 2} \omega_0,
 \ee
it is independent of time.

A quick way to derive these results is to work with the shifted
creation and destruction operators,
\be
\label{shiftop}
\tilde{a}=a - \alpha, \tilde{a}^\dagger=a^\dagger - \alpha \ee where
$\alpha$ is given in eq.(\ref{valpara}). The Hamiltonian takes the
form, \be
\label{newh}
H=\omega_0(\tilde{a}^\dagger \tilde{a}) + {1\over 2} \omega_0
\ee
It is clear then that the ground state is annihilated
by $\tilde{a}$, leading to
eq.(\ref{agndstate}) and the ground state energy is eq.(\ref{gndstateen}).

 For the quantum adiabatic theorem to be valid, the condition in
 eq.(\ref{condada}) must hold. For the harmonic oscillator it is easy to
 see that this gives, \be
 \label{qadsho}
 \dot{J} \ll \omega_0^{5/2}.
 \ee

 In fact the time evolution in this case can be exactly solved.  We
 consider the case where $J(t)\rightarrow 0, t\rightarrow -\infty$.
 Starting with the state $|0>$ in the far past, which is the vacuum of
 the Hamiltonian in the far past, we then find that the state at any
 time $t$ is given by, \be
 \label{statet}
 |\psi(t)>=N(t) e^{\lambda(t) a^\dagger}|\phi_0> \ee
where $|\phi_0>$
 is the adiabatic vaccum given in eq.(\ref{agndstate}), $N(t)$ is a
 normalisation constant and the coherent state parameter is
 $\lambda(t)$. Imposing Schrodinger equation one gets
\be
i{\dot{\lambda}}  =  i\frac{\dot{J}}{\sqrt{2\omega_0^3}}+\omega_0
\lambda.
\label{coherenteqns}
\ee
The solution for $\lambda(t)$ with initial condition $\lambda(-\infty)
= 0$ is given by,
\be
 \label{cstatepara}
 \lambda(t)={e^{-i\omega_0t}\over \sqrt{2\omega_0^3}}\int_{-\infty}^t
 \dot{J}(t') e^{i\omega_0t'}dt'.
\ee
Some details leading to
 eq. (\ref{coherenteqns})  are given in Appendix \ref{appb}.
This state will
 behave like a classical state when the coherent state parameter is
 big in magnitude, i.e., when \be
 \label{condclassicala}
 |\lambda|\gg 1.
 \ee

 The integral on the rhs of eq.(\ref{cstatepara}) can be done by parts (we set
 $J(-\infty)=0$),
 \be
 \label{ct2}
 \int_{-\infty}^t dt^\prime~
\dot{J}e^{i\omega_0t'}=\dot{J}(t){e^{i\omega_0t}\over
 i\omega_0}-\int_{-\infty}^t dt' \ddot{J} {e^{i\omega_0t'}\over
 i\omega_0}. \ee Subsequent iterations obtained by further integrations
 by parts gives rise to a series expansion\footnote{In general one expects this to be an
 asymptotic rather than convergent series.}  for $\lambda$ in terms of
 higher derivatives of $J$. The higher order terms are small if $J$ is
 slowly varying compared to the frequency of the oscillator
 $\omega_0$. Evaluating the second term which arises in his expansion
 for example and requiring it to be smaller than the first term in
 eq.(\ref{ct2}) gives,
\be
 \label{correccpa}
 {\ddot{J}\over \dot{J} \omega_0}\ll 1
\ee
We assume now that $J$ is
 slowly varying and the first term on the rhs of eq.(\ref{ct2}) is a
 good approximation to the integral.
This tells us that for
 eq.(\ref{condclassicala}) to be true the condition which must be met
 is, \be
 \label{condclassc}
 \dot{J}\gg \omega_0^{5/2}.  \ee Note that this condition is opposite
 to the one needed for the quantum adiabatic theorem to apply eq.(\ref{qadsho}).

 The answer for the $<X>$ can be easily obtained by inserting the expression
 for $\lambda$ obtained in  eq.(\ref{cstatepara}) in the wave function,
 eq.(\ref{statet}). Let us obtain it here in a slightly different
 manner. When eq.(\ref{condclassicala}) is true the system behaves
 classically and its response to the driving force can be obtained by
 solving the classical equation of motion for the forced
 oscillator. In terms of the fourier transform of $J$ this gives, \be
 \label{famans}
 X(t)=\int {J(\omega)\over \omega^2-\omega_0^2}e^{-i\omega t } d\omega
 \ee The correct pole prescription on the rhs is that for a retarded
 propagator.

 When the source is slowly varying compared to
 $\omega_0$, the denominator $\omega^2-\omega_0^2$ in
 eq.(\ref{famans}) can be expanded in a power series in $\omega^2\over
 \omega_0^2$ and the resulting fourier transforms can be expressed as
 time derivatives of $J$. The first two terms give, \be
\label{ftterm}
X=-{J(t)\over \omega_0^2}+{\ddot{J}\over \omega_0^4}+ \cdots \ee The
first term on the rhs is the location of the instantaneous
minimum. The second term is the first correction due to the time
dependent source. Subsequent corrections are small if the source is
slowly varying and condition eq.(\ref{correccpa}) is met. It is useful
to express this result as, \be
\label{ftthree}
X+{J(t)\over \omega_0^2}={\ddot{J}\over \omega_0^4}+ \cdots.
\ee The
left hand side is  the expectation value of $X$ after adding a shift to account for
the  instantaneous minimum of the potential. The right hand side we see
now only contains time derivatives of $J$. Before proceeding let us note that the
 expanding the denominator in eq.(\ref{famans}) in a power series in
${\omega^2\over \omega_0^2}$ gives  a good approximation only if
$J(\omega)$ has most of its support for $\omega \ll \omega_0$.
 This is how the more
 precise condition  mentioned in the footnote before eq.(\ref{oscconda})
arises.

It is also useful to discuss the energy.  From eq.(\ref{ftterm}) and
the Hamiltonian we see that the leading contribution comes from the
Kinetic energy term and is given to leading order by, \be
\label{en}
E={1\over 2}{\dot{J}^2\over \omega_0^4}
\ee
(strictly speaking this is the energy above the ground state energy).

The external source driving the oscillator changes its
energy. Noether's argument in the presence of the time dependent
source leads to the conclusion that \be
\label{nar}
{ \partial H \over \partial t}=\dot{J} (X+{J\over \omega_0^2}) \ee (this
also directly follow from the Hamiltonian, eq.(\ref{tdef})). From
eq.(\ref{ftthree}) and eq.(\ref{en}) we see that this condition is
indeed true.  Let us also note that the rate of change in energy can
be expressed in terms of the shifted operators, eq.(\ref{shiftop}),
as, \be
\label{cshift}
{\partial H \over \partial t}=\dot{J}
({\tilde{a}+\tilde{a}^\dagger\over \sqrt{2\omega_0}}), \ee this form will
be useful in our discussion below.

To summarise, we find that when the conditions eq.(\ref{condclassc}),
 eq.(\ref{correccpa}), are met the driven harmonic oscillator behaves
 like a classical system. Its response, for example, $<X>$, and the
 energy, $E$, can be calculated in an expansion in time derivatives of
 $J$, which is controlled when eq.(\ref{correccpa}) is valid and the
 source is slowly varying. We will refer to this perturbation
 expansion as the classical adiabatic perturbation approximation
 below. Note that the condition, eq.(\ref{condclassc}) is opposite to
 the one required for the quantum adiabatic perturbation theory to
 hold. In the next subsection we will discuss how a similar classical
 adiabatic approximation arises in the gauge theory.

\section{Gauge Theory: Large N Classical
Adiabatic Perturbation Theory (LNCAPT)}

We now return to the gauge theory and formulate a large $N$ classical
adiabatic approximation based on coherent states in this theory. This
will allow us to obtain results in the gauge theory which agree with
those obtained using supergravity in \S2, \S3.

\subsection{Adiabatic Approximation in terms of Coherent States}

The supergravity solution in \S2 describes {\em classical solutions}
rather than states which contain a small number of bulk particles. The
AdS/CFT correspondence implies that bulk classical solutions
corresponds to coherent states in the boundary gauge theory with a large number of particles in which
operators like $\hat{{\cal O}}$ have nontrivial expectation values. On the
other hand, states obtained by the action of a few factors of $\hat{{\cal O}}$
on the vaccum are few-particle states in the bulk. The quantum adiabatic
approximation described in \S4 attempts to determine the wave function
in a basis formed out of such single particle states and does not apply to the supergravity solution in \S2.

We, therefore, need to formulate an adiabatic approximation in terms
of coherent states of gauge invariant operators in the boundary theory
to try and understand the supergravity solutions of \S2 in a dual
description. As is well known, these coherent states become classical
in a smooth fashion in the $N \rightarrow \infty$ limit. (See
e.g.~\cite{Yaffe:1981vf}).  Consider a complete (usually overcomplete)
set of gauge invariant operators in the Schrodinger picture, $\hat{{\cal O}}^I$. A general coherent state is of the form \ben |\Psi (t)> =
{\rm exp} \left[i\chi (t) + {\sum_I \lambda^I (t)
\hat{{\cal O}}^I_{(+)}}\right] |0>_A.
\label{cone}
\een
Here $\hat{{\cal O}}^I_{(+)}$ denotes the creation part of the operator and  $|0>_A$ denotes the adiabatic vacuum corresponding to some
instantaneous value of the dilation $\Phi_0$,
\ben
H [\Phi_0] |0>_A = E_{\Phi_0} |0>_A
\label{ccone}
\een
with the ground state energy $E_{\Phi_0}$.

The algebra of operators
$\hat{{\cal O}}^I$, together with the Schrodinger equation then leads to a
differential equation which determines the time evolution of the
coherent state parameters $\lambda^I (t)$ in terms of the time
dependent source $\Phi_0 (t)$. The idea is then to solve this equation
in an expansion in time derivatives of $\Phi_0 (t)$. This is the
coherent state adiabatic approximation we are seeking.

In general it is almost impossible to implement this program
practically,
since the operators
$\hat{\mathcal{O}}^I$ have a non-trivial operator algebra which mixes all of
them. The coherent state (\ref{cone}) is in the co-adjoint orbit of
this algebra \cite{Yaffe:1981vf}. The resulting theory of fields
conjugate to these operators would be in fact the full interacting
string field theory in the bulk. In our case, however, the situation
drastically simplifies for large 't Hooft coupling
at the lowest order of an expansion in ${\dot
\Phi}_0$. This is because these various operators decouple and their
algebra essentially reduces to free oscillator algebras.

We have already found this decoupling in our supergravity
calculation. The departure of the solution from $AdS_5 \times S^5$ is
due to the time-dependence of the boundary value of the dilaton, and
are small when the time variations are small, controlled by the
parameter $\epsilon$. To lowest order in $\epsilon$ (which is
$O(\epsilon^2)$) the deformation of the bulk dilaton in fact satisfied
a linear equation in the $AdS_5$ background in the presence of a
source provided by the boundary value $\Phi_0(t)$. This equation does
not involve the deformation of the metric. Similarly, the equation for
metric deformation does not involve the dilaton deformation to lowest
order.

This allows us to treat each supergravity field and its dual
operator separately. With this
understanding we will now consider the coherent state (\ref{cone})
with only the operator dual to the dilaton, $\hat{{\cal O}}$. Since our source
is spherically symmetric and higher point functions of the operators
are not important in this lowest order calculation, we can restrict
this operator to its spherically symmetric part.

\subsection{Large N Classical Adiabatic Perturbation Theory (LNCAPT)}
Let us now elaborate in more detail on the LNCAPT.

The linearised approximation in the gravity theory means that only the
two point function is non-trivial and all connected higher point
functions vanish. The non-linear
terms correspond to nontrivial higher order correlations.
In this approximation
the gauge theory simplifies a great deal.  Each gauge invariant
operator- which is dual to a bulk mode- gives rise to a tower of
harmonic oscillators. The response of the gauge theory can be
understood from the response of these oscillators.

In fact in the quadratic approximation the only oscillators which are
excited are those which couple directly to the dilaton and so we only
have to discuss their dynamics.  We have already discussed the
operator dual to the dilaton in section \S4.3. The dilaton excitations
we consider are $S^3$ symmetric and correspondingly the only modes of
$\hat{{\cal O}}$ which are excited are $S^3$ symmetric. Here we denote
these by $\hat{{\cal O}}_{l=0}$.

In the Heisenberg picture $\hat{{\cal O}}_{l=0}$ can be expanded in
terms of time dependent modes, this is dual to the fact that the $S^3$
symmetric dilaton can be expanded in terms of modes with different
radial and related time dependence in the bulk.  One finds, as is
discussed in Appendix \ref{normal}, that only even integer frequencies
appear in the time dependence giving, \be
\label{hpic}
\hat{{\cal O}}_{l=0}=N \sum_{n=1}^\infty F(2n)[A_{2n} e^{-i2n
t}+A^\dagger_{2n} e^{i2nt}]. \ee
Here $A_{2n}, A_{2n}^\dagger$ are
canonically normalised creation and destruction operators satisfying
the relations, \ben [ A_m , A_n ] = [ A^\dagger_m , A^\dagger_n ] = 0
~~~~~~~ [ A_m , A^\dagger_n ] = \delta_{m,n}.
\label{cthree}
\een
Their commutators
with the gauge theory hamiltonian are
\ben
[ H , A^\dagger_{2n} ] = (2n) A^\dagger_{2n}~~~~~~~~~~~~~~~~~~~~
[ H , A_{2n} ] = - (2n) A_{2n}
\label{cseven}
\een
The normalization factor $F(2n)$ may be computed by comparing
with the standard the 2-point function as is detailed in Appendix
\ref{normal}. The result is \ben |F(2n)|^2 = \frac{A
\pi^4}{3}~n^2 (n^2-1)  
\label{csix}
\een
for $n\ge 2$. $F(0)$ and $F(2)$ vanish, so this means that the
sum in eq.(\ref{hpic}) receives its first contribution at $n=2$.  It
also means that the lowest energy state which can be created by acting
with $\hat{{\cal O}}_{l=0}$ on the vacuum has energy equal to $4$.
This is what we expect on general grounds, since the energies of
states created by an operator with conformal dimension $\Delta$ are
given by \ben \omega (n,l) = \Delta + 2n + l(l+2)~~~~~~~n=0,1,2\cdots
\label{cthreea}
\een

The constant $A$ in eq.(\ref{csix}) is the normalization of the
2-point function which may be determined e.g. from a bulk
calculation. Before proceeding let us also note that $F(2n)$ grows like
$F(2n) \sim n^2 $, eq.(\ref{csix}), for large mode number $n$. This
enhances the coupling of the higher frequency modes to the dilaton and
will be important in our discussion of renormalisation below.

From now onwards we will find it convenient to work in the Schrodinger
representation, in which operators are time independent. The operator
$\hat{{\cal O}}_{l=0}$ in this representation is given by, \be
\label{ophatosr}
\hat{{\cal O}}_{l=0}= N \sum_n F(2n) [A_{2n} + A_{2n}^\dagger].
\ee

From eq.(\ref{cseven}) it follows that the Hamiltonian for $A_{2n},
A_{2n}^\dagger$ modes can be written as,
\be
\label{hamdilmodes}
H=\sum_n 2n A_{2n}^\dagger A_{2n}.
\ee
Note this Hamiltonian measures the energy above that of the ground state.

The operators, $A_{2n}^\dagger , A_{2n}$  create and destroy a single
  quantum of excitation when acting on the vaccum of the ${\cal N}=4$
  theory with the instantaneous value of
  $g^2_{YM}=e^{\Phi_0}$. Thus they are the analogue of the shifted
  creation and destruction operators we had defined in the harmonic
  oscillator case, $\tilde{a}, \tilde{a}^\dagger$. The Hamiltonian,
  eq.(\ref{hamdilmodes}), is the analogue of the Hamiltonian,
  eq.(\ref{newh}) in the harmonic oscillator case.

The time dependence of the Hamiltonian due to the varying dilaton can
be expressed as follows,
\be
\label{timedepc8}
{\partial H \over \partial t}={\partial H \over \partial \Phi}
\dot{\Phi}_0=-\hat{{\cal O}}_{l=0}\dot{\Phi}_0
\ee
leading to,
\be
\label{timedepc9}
{\partial H \over \partial t}=-\hat{{\cal O}}_{l=0} \dot{\Phi}_0=-N \sum_n
F(2n)[A_{2n}+A_{2n}^\dagger] \dot{\Phi}_0,
\ee
where we have used
eq.(\ref{ophatosr}).  It is useful to write this as \be
\label{timedepc10}
{\partial H \over \partial t}= -N \sum F(2n)\sqrt{4n} \dot{\Phi}_0
[{A_{2n}+A_{2n}^\dagger \over \sqrt{4n}}],
\ee
which is analogous to the
time dependence in the forced oscillator system, eq.(\ref{cshift}).

So we see that the gauge theory, in the quadratic approximation maps
to a tower of oscillators, with frequencies, $\omega_n=2n$.
Comparing with eq.(\ref{cshift}) we see that the oscillator with
energy $2n$ couples to a source,
\be
\label{ds}
\dot{J_n}= -N F(2n) \sqrt{4n} \dot{\Phi}_0.
\ee

The analysis of the harmonic oscillator now directly applies.  The
resulting state is a coherent state, \be
\label{csgt}
|\psi>=\hat{N}(t) e^{(\sum_n \lambda_n A_{2n}^\dagger)} |\phi_0>.
\ee
Here
$|\phi_0> $ is the adiabatic vacuum, which in is the ground state of
the ${\cal N}=4$ theory with coupling $g_{YM}^2=e^{\Phi_0}$.  $\hat{N}(t)$ is a normalisation
constant and the  coherent
state parameter $\lambda_n$ is given from eq.(\ref{cstatepara}) by,
\be
\label{valln}
\lambda_n={e^{-i\omega_n t}\over {\sqrt{2 \omega_n^3}}}
 \int_{-\infty}^t \dot{J}_n(t')
e^{i\omega_n t'} dt'.
\ee

The condition that the source is varying slowly, eq.(\ref{correccpa}), becomes,
\be
\label{condslowvara}
|{\ddot{\Phi}_0 \over n \dot{\Phi}_0}|  \ll 1~~~~~~\forall n.
\ee
It is clearly sufficient to satisfy this condition for $n=1$,
\be
\label{condslowvar}
|{\ddot{\Phi}_0 \over \dot{\Phi}_0}| \sim \epsilon  \ll 1.
\ee

 This
condition is met for the dilaton profile we have under consideration \footnote{This condition
is analogous to eq.(\ref{oscconda}) for the driven harmonic oscillator. As discussed
in that context in the footnote before eq.(\ref{oscconda}) there is a more precise version
of this condition. It is the statement that for all modes, $n$,
the fourier transform of $J_n$ must have essentially all its support at
  frequencies  much
smaller than the oscillator frequency, $2n$.}.
When this condition is true $\lambda_n$ can be evaluated by keeping
the first term in eq.(\ref{ct2}). The condition that the state is
classical, is that $\lambda_n \gg 1$, this gives\footnote{The more precise condition
is simply that $\lambda_n \gg 1, \ \  \forall n$. This gives, eq.(\ref{condcln}) provided
that the integral in eq.(\ref{valln}) can be approximated by the first  term of the
derivative approximation.},
\be
\label{condcln}
|N F(2n) \sqrt{4n} \dot{\Phi}_0| \gg (2n)^{5/2}.
\ee

Noting from eq.(\ref{csix}) that $F(2n) \sim n^2$ for large $n$ we see that the factors of $n$ cancell out on both sides, leading to the conclusion that when,
\be
\label{condeps}
|N \dot{\Phi}_0| \sim N \epsilon \gg 1
\ee
all the oscillators are in a classical state.
In this way we recover the first condition discussed in eq.(\ref{condcap1}).

The summary is that when the two conditions,
\be
\label{twocond}
\epsilon \ll 1, N \epsilon \gg 1
\ee
are both valid, the gauge theory
is described to leading order in $\epsilon$ as a system of harmonic
oscillators. The oscillators which couple to the dilaton are excited
by it and are in a classical state.

This description can be used to calculate the resulting expectation
value of operators. The calculation for $<{A_{2n}+A_{2n}^\dagger \over
\sqrt{4n}}>$ is analogous to that for $<X+{J\over \omega^2}>$ in the
harmonic oscillator case (since the $A_{2n}, A_{2n}^\dagger$ are analogous
to the shifted operators, $\tilde{a}, \tilde{a}^\dagger$
eq.(\ref{shiftop})).  From eq.(\ref{ftthree}) and eq.(\ref{ds}) we get
that to leading order in $\epsilon$,
\be
\label{reshato}
<{A_{2n}+A_{2n}^\dagger \over \sqrt{4n}}>=-N {F(2n)\sqrt{4n}\over (2n)^4}
\ddot{\Phi}_0.
\ee
Substituting in eq.(\ref{ophatosr}) next gives,
\be
\label{reshato2}
<\hat{{\cal O}}_{l=0}>= -C N^2 \ddot{\Phi}_0
\ee
where $C$ is
\be
\label{defCC}
C= \sum {F(2n)^2\over 4n^3}.  \ee
The functional dependence on $\Phi_0$
and $N$ in eq.(\ref{reshato}) agrees with what we found in the
supergravity calculation, eq.(\ref{exdilf}).  The constant of
proportionality $C$ is in fact quadratically divergent. This follows
from noting that for large $n$, $F(2n)\sim n^2$.

A little thought tells us that the divergence should in fact have been
expected. The supergravity calculation also had a divergence and the
finite answer in eq.(\ref{exdilf}) was obtained only after regulating
this divergence and renormalising. Therefore it is only to be expected
that a similar divergence will also appear in the description in terms
of the oscillators.  In the subsection which follows we will discuss
the issue of renormalisation in more detail. The bottom line is that
counter terms can be chosen so that the coefficient in
eq.(\ref{exdilf}) agrees with that in the supergravity calculation.

It is also important to discuss how the energy behaves.  From
eq.(\ref{en}) and eq.(\ref{ds}) we see that the energy above the
ground state is
\be
\label{enabgnd}
<E>-E_{gnd}={1\over 2} C N^2 \dot{\Phi}_0^2
\ee
We note that the functional dependence on $\dot{\Phi}_0, N$ match with
those obtained in the supergravity calculations, eq.(\ref{fmass}). The
constant of proportionality which is obtained by summing over the
oscillator modes in the case of the energy is the same as $C$ defined
above, eq.(\ref{defCC}). It is also therefore quadratically divergent.

The fact that the two constants of proportionality in
eq.(\ref{enabgnd}) and eq.(\ref{reshato2}) are the same follows on
general grounds. Noether's argument in the presence of the time
dependence means that each oscillator satisfies the relation,
eq.(\ref{nar}).  On summing over all of them we then get the relation
\be
\label{condeq}
<{dE\over dt}>=-\dot{\Phi}_0 <\hat{{\cal O }}_{l=0}>
\ee
leading to the equality of the two constants. Earlier we had also seen
that the supergravity calculation satisfies this relation,
eq.(\ref{noether}). It follows from these observations that if after
renormalisation the answer for $<\hat{{\cal O }}_{l=0}>$ agrees between the
supergravity theory and the oscillator description developed here, then the
expectation value for $E$ will also agree in the two cases.

Here we have analysed the gauge theory to leading order in $\epsilon$.
Going to higher orders introduces anharmonic couplings between the
different oscillators. These couplings arise because of connected
three-point and higher point correlations in the gauge theory. The
three point function for example is suppressed by $1/N$, the four
point function by $1/N^2$ and so on. For computations in the ground
state these would therefore be suppressed in the large $N$
limit. However as we have seen here the time dependence results in a
coherent state which contains $O(N \epsilon)^2$ quanta being
produced. The $3$- pt function in such a state is suppressed by
$O(\epsilon)$ and not by $O(1/N)$. Since $\epsilon \ll 1$, this
is still enough though to justify our neglect of the cubic terms to leading
order in $\epsilon$. Similarly the effect of $4$-pt correlators in the
coherent state are suppressed by $O(\epsilon)^2$ etc.  This is in
agreement with the supergravity calculation, where the cubic terms in the
equations of motion are suppressed by $O(\epsilon)$ etc.

To go to higher orders in $\epsilon$ using the oscillator description
the effect of the anharmonic couplings induced by the higher order
correlations would have to be introduced. In addition one would have
to keep the contributions from the quadratic approximation to the required
order in $\epsilon$.  As long as the 'tHooft coupling stays big for
all times and the supergravity approximation is valid, there is no reason to
believe that these effects will be significant and the behaviour of
the system should be well described by the leading harmonic oscillator
description, in agreement with what we saw in supergravity. When the 'tHooft
coupling begins to get small though the anharmonic couplings could
potentially significantly change the behaviour of the system, as we
will discuss in section 6.4.

\subsection{Renormalisation}

Let us now return to the constant $C$ eq.(\ref{defCC}). One would like
to know if it can be made to agree with the supergravity answer
eq.(\ref{exdilf}). Since the mode sum in $C$ diverges, at first sight
it would seem that by suitably removing the infinities this can always
be done. To be explicit, imposing a cutoff on the mode sum in $C$ one
gets from eq.(\ref{defCC}),
\be
\label{mmodesum}
C= \sum {F(2n)^2\over 4n^3}=c_1 n_{max}^2 + c_2 \ln (n_{max}) + {\rm
finite~ term}
\ee
(A term linear in $n_{max}$ can always be removed by
shifting $n_{max}$).  Removing the infinities would mean removing the
first two terms, but by changing $n_{max}$ by a finite amount the
finite term left over will clearly change and can be made equal to any
answer we want.

However this seems too superficial an answer. One would like to ensure
that the freedom to adjust $C$ corresponds to the freedom to add local
counterterms in the theory, and also that once the counter terms are
chosen so that $C$ agrees no other discrepancy appears with supergravity.

This is in fact true and can be easily seen by relating the
calculation for $<\hat{{\cal O}}>$ in eq.(\ref{reshato2}) to the
two-point function for the dilaton.  In fact we will only need the two
point function of the S-wave dilaton which is equal to the two-point
function of $<\hat{{\cal O}}_{l=0}\hat{{\cal O}}_{l=0}>$ in the gauge
theory. Since the S-wave dilaton couples directly to $
\hat{{\cal O}}_{l=0}$, we have
\ben
< \hat{{\cal O}}_{l=0} (t) > =
\int dt^\prime~<\hat{{\cal O}}_{l=0}(t)\hat{{\cal O}}_{l=0}(t') >
\Phi (t^\prime)
\label{onetwopoint}
\een
Using eq.(\ref{hpic}) we find that
\be
\label{tpt}
<\hat{{\cal O}}_{l=0}(t)\hat{{\cal O}}_{l=0}(t') >
=N^2 \sum_n F(2n)^2 (4n) \int {d\omega\over 2 \pi i}
{e^{-i(t-t')\omega}\over (\omega^2-(2n)^2)}
\ee
where we have expressed the answer in terms of a fourier transform in
frequency space.  We are not being explicit about the pole
prescription here, this will determine which propagator (Feynman,
Retarded etc) one requires.  From eq.(\ref{tpt}) the propagator in
frequency space can be read off to be, \be
\label{propfr}
G(\omega)=N^2\sum_n {F(2n)^2 (4n)\over (\omega^2-(2n)^2)}
\ee
Since $F(2n) \sim n^2$ the sum over modes on the rhs is {\it
quartically} divergent.

For purposes of comparing with the adiabatic approximation we expand
this propagator in powers in $\omega^2$. This gives,
\be
\label{expo}
{G(\omega)\over N^2}=-\sum {F(2n)^2 (4n) \over (2n)^2} - \omega^2 \sum
{F(2n)^2 (4n) \over ((2n)^2)^2}- \omega^4 \sum {F(2n)^2 (4n) \over
((2n)^2)^3} + \cdots
\ee
The terms within the ellipses contain powers
higher than $\omega^4$ and are not divergent. The first term on the
rhs must be set to zero after renormalisation to preserve conformal
invariance, otherwise the vacuum expectation value for $<\hat{{\cal
O}}>$ in the ${\cal N}=4$ theory with constant coupling would not
vanish.  The leading contribution to $<\hat{{\cal O}}>$ in the
adiabatic approximation then arises from the second term which is
quadratically divergent.  After fourier transforming the $\omega^2$
dependence of this term gives rise to the second derivative with
respect to the time of the dilaton. And the sum over modes is the same as that in $C$,
eq.(\ref{defCC}).

Now the point is that all divergences in the two-point function can be
 removed by local counterterms since they correspond to contact
 terms. In fact the gravity calculation also needed counterterms and
 from our discussion in \S3.1 we know that these counterterms are of
 the form given in eq.(\ref{cterm}). In particular the third term in
 eq.(\ref{cterm}) proportional to $(\nabla\Phi)^2$ cancels the
 quadratic divergence while the last term in eq.(\ref{cterm}),
 $a_{(4)}$, contains terms which cancel the subleading logarithmic
 divergence.  Also once the counter terms are chosen so that $C$
 agrees no other discrepancy can appear. The point here is that the
 leading order in $\epsilon$ calculations are only sensitive to the
 two-point function. And the finite terms in the two-point function
 are well known to agree between the gravity and gauge theory sides.
 In fact the finite two point function is just determined by conformal
 invariance and since the anomalous dimension of $\hat{{\cal O}}$ does
 not get renormalised, it can be calculated in the free field limit
 itself.

The bottom line then is that using the freedom to adjust the counter
terms, $C$ can be made to agree with the supergravity calculations in \S3.

Let us end by pointing out that the supergravity value for $C$,
eq.(\ref{exdilf}) is, \be
\label{suval}
C_{sugra}={1\over 16}
\ee
which means that the effect of renormalisation is to only include the
contributions of modes with mode number $n\sim O(1)$. This makes good
physical sense, we are dealing with the low frequency response of the
system here, and the high frequency modes should not be relevant for
this purpose.

This last comment also has a bearing on our discussion in \S4 of the
 quantum adiabatic perturbation theory. The criterion for the validity
 of this approximation was stated in eq.(\ref{condrz}). Now what this
 condition really ensures is that the amplitude to excite the system
 to a state $|\phi_n>=A_n^\dagger |0>$ containing any one single
 oscillator excitation is small.  However there are an infinite number
 of such single excitation states, corresponding to the infinite
 number of values that $n$ takes, and one might be worried that this
 condition is not sufficient. Even though the amplitude to excite the
 system into any given state $|\phi_n>$ is small the sum of these
 amplitudes, more correctly the norm of the first order correction of
 the wave function $<\psi^1|\psi^1>$, eq.(\ref{fcorr}), is still be
 large and in fact would diverge when summed over all the modes. This
 would invalidate the approximation. The reason this concern does not
 arise is tied to our discussion above. After renormalisation only a
 few low frequency modes contribute to the response of the system and
 one is only interested in how the wave function changes for these
 modes. For this purpose the condition in eq.(\ref{condrz}) is enough
 and we see that when it is met the quantum adiabatic approximation is
 indeed valid.

\subsection{Highly Curved Geometry}
So far we have considered what happens in the parametric regime,
eq.(\ref{twocond}), when the 'tHooft coupling stays big all times. In
this case the supergravity description is always valid.  We saw above that
the gauge theory can be described in this regime in terms of
approximately decoupled classical harmonic oscillators and this
reproduces the supergravity results.

Now let us consider what happens when the dilaton takes a larger
excursion so that the 'tHooft coupling at intermediate times becomes
of order unity or even smaller.  Some of the resulting discussion is
already contained in the introduction above.

A natural expectation is that description in terms of classical
adiabatic system of weakly coupled oscillators should continue to
apply even when the 'tHooft coupling becomes small.  There are several
reasons to believe this. First, anharmonic terms continue to be of
order $\epsilon$ and thus are small. The leading anharmonic terms
arise from three -point correlations,
$<\hat{O}_1\hat{O}_2\hat{O}_3>$. In the vaccum these go like $1/N$. In
the coherent state produced by the time dependence these go like
$\epsilon$.
The enhancement by $N \epsilon$ arises
because the coherent state contains $O((N \epsilon)^2)$ quanta, so
that the probability goes as $(N\epsilon)^2/N^2 \sim \epsilon^2$
\footnote{The probability 
\mbox{$|<\phi|\hat{{\cal O}}\hat{{\cal O}}\hat{{\cal O}}|\phi>|^2$} is
proportional to $\frac{1}{N^2}(N^2\epsilon^2)^3$, with each factor of
$N^2\epsilon^2$ as an estimate of the contribution for each of the
operators $\hat{{\cal O}}$.  The contribution of the $2$-pt function
\mbox{$|<\phi|\hat{{\cal O}}\hat{{\cal
O}}|\phi>|^2$} is just proportional to $(N^2\epsilon^2)^2$, resulting
in a relative suppression of $O(\epsilon^2)$.}.
Four-point
functions give rise to terms going like $O(\epsilon^2)$ and so on,
these are even smaller.  In the absence of anharmonic terms the theory
should reduce to a system of oscillators. Second, the existence of a
gap of order $1/R$ means that for each oscillator the time dependence
is slow compared to its frequency. Therefore the system continues to
be very far from resonance and should evolve adiabatically. Finally,
in the parametric regime, eq.(\ref{twocond}) the analysis of the
previous subsections should then apply leading to the conclusion that
an $O(N \epsilon)\gg 1$ quanta are produced making the coherent state
a good classical state.

If this expectation is borne out the system should settle back into
the ground state of the final ${\cal N}=4$ theory in the far future
and should have a good description in terms of smooth AdS space then.

However, as discussed in the introduction, there are reasons to worry
that this expectation is not borne out. New features could enter the
dynamics when the 'tHooft coupling becomes small at intermediate
times, and these could change the qualitative behaviour of the
system. These new features have to do with the fact that string modes
can start getting excited in the bulk when the curvature becomes of
order the string scale. These modes correspond to non-chiral operators
in the gauge theory and the corresponding oscillators have a time
dependent frequency. When the 'tHooft coupling is big these
frequencies are much bigger than those of the supergravity modes and
as a result the string mode oscillators are not excited. But when the
'tHooft coupling becomes of order unity some of the frequencies of
these string modes become of order the supergravity modes and hence
these oscillators can begin to get excited \footnote{The primary
reason for them getting excited are the anharmonic terms which couple
them to the modes dual to the dilaton.}.  In fact the string modes are
many more in number than the supergravity modes, since there are an order
unity worth of chiral operators in the gauge theory and an $O(N^2)$
worth of non-chiral ones.

The worry then is that if a significant fraction of these string
oscillators get excited the correct picture which could describe the
ensuing dynamics is one of thermalisation rather than classical
adiabatic evolution. In this case the energy pumped into the system
initially would get equipartitioned among all the different degrees of
freedom.  Subsequent evolution would then be dissipative, and the
energy would increases in a monotonic manner, as it does for a large
black hole, eq.(\ref{tempeq}).

Due to the dissipative behaviour the energy which is initially pumped
in would not be recovered in the future. Rather one would expect that
when the 'thooft coupling becomes large again, the energy, which is of
order $N^2\epsilon^2$ remains in the system. The gravity description
of the resulting thermalized state depends on the value of $\epsilon$
relative to $ \lambda\equiv g_{YM}^2 N$ and $ N$. In this late time regime of large 't
Hooft coupling, the various possibilities can be figured out from
entropic considerations in supergravity ( see e.g. section 3.4 of
\cite{Magoo}). The result in our case is the following.
For $\epsilon \ll (g_{YM}^2N)^{5/4}/N$ a gas of supergravity modes is
favored. For $(g_{YM}^2N)^{5/4}/N < \epsilon \ll (g_{YM}^2 N)^{-7/8}$ one
would have a gas of massive string modes. For $(g_{YM}^2N)^{-7/8} <
\epsilon \ll 1$ one gets a small black hole, i.e. a black hole whose
size is much smaller than $R_{AdS}$. A
big black hole requires $O(N^2)$ energy which is
parametrically much larger. Thus, the strongest departure from $AdS$
space-time in the far future would be presence of small black holes.
Such black holes would eventually evaporate by
emitting Hawking radiation. However this takes an $O(N^2 R_{AdS})$
amount of time which is much longer than the time scale
$O(R_{AdS}/\epsilon)$ on which the 'tHooft coupling evolves. As a result for
a long time after the 'tHooft coupling has become big again the
gravity description would be that of a small black hole in AdS space.

An important complication in deciding between these two possibilities
 is that the rate of time variation is $\epsilon$ which is also the
 strength of the anharmonic couplings between the supergravity oscillators
 and string oscillators.  If the rate of time variation could have
 been made much smaller, thermodynamics would become a good guide for
 how the system evolves. In the microcanonical ensemble, which is the
 correct one to use for our purpose, with energy $N^2\epsilon^2$ the
 entropically dominant configurations are as discussed in the previous
 paragraph, and this
 would suggest that dissipation would indeed set in. However, as
 emphasised above this conclusion is far from obvious here since the
 time variation is parametrically identical to the strength of the
 anharmonic couplings.

In fact we know that the guidance from thermodynamics is misleading in
 the supergravity regime, where the 'tHooft coupling stays large for all
 times.  In this case we have explicitly found the solution in \S2. It
 does not contain a black hole. Moreover, it does not suffer from any
 tachyonic instability - since it is a small correction from AdS space
 which does not have any tachyonic instability \footnote{Note that we
 are working on $S^3$ here.}.  The only way a black hole could form is
 due to a tunneling process but this would be highly suppressed in the
 supergravity regime.

One reason for this suppression is that the energy in the supergravity
 solution discussed in \S2 is carried by supergravity quanta which have a
 size of order $R_{AdS}$.  This energy would have to be concentrated
 in much smaller region of order the small black hole's horizon to
 form the black hole and this is difficult to do.  In contrast, away
 from the supergravity regime this could happen more easily. When the 'tHooft
 coupling becomes small at intermediate times, strings become large
 and floppy, of order $R_{AdS}$, at intermediate times.  If a
 significant fraction of the energy gets transferred to these strings
 at intermediate times it could find itself concentrated within a
 small black hole horizon once the 'tHooft coupling becomes large
 again.

In summary we do not have a clean conclusion for the future fate of
 the system in the parametric regime, eq.(\ref{twocond}). Note however
 that in both possibilities discussed above most of space-time in the
 far future is smooth $AdS$ space, with the possible presence of a
 small black hole.  Hopefully, the framework developed here will be
 useful to think about this issue further.

\section{Conclusions}

In this paper we examined the behaviour of the $AdS_5\times S^5$
solution of IIB supergravity when it is subjected to a time dependent
boundary dilaton. This is dual to the behaviour of the ${\cal N}=4$
Super Yang -Mills theory subjected to a time dependent gauge
coupling. The $AdS_5$ solution was studied in global coordinates and
the dual field theory lives on an $S^3$ of fixed radius $R$. We worked
in units where $R_{AdS}=R=1$.  Three parameters are relevant for
describing the resulting dynamics:

\begin{enumerate}

\item{} $N$ - which is the number of units of flux and is dual to the rank
of the gauge group. This was held fixed during the evolution.

\item{} $\lambda = e^{\Phi(t)} N$ - which determines the value of $R_{AdS}$
in string units is the 'tHooft coupling in the gauge
theory. Especially relevant is its minimum value $\lambda_{min}$
during the time evolution.  When $\lambda_{min}\gg 1$ supergravity is
a good approximation for all times. When $\lambda_{min}\le O(1)$ supergravity
breaks down at intermediate times.

\item{} $\epsilon \sim \dot{\Phi}$ - which determines the rate of change of
 the boundary dilaton in units of $R_{AdS}$.  Throughout the analysis
 we worked in the slowly varying regime where $\epsilon \ll 1$.

\end{enumerate}

Our results are as follows:

\begin{itemize}

\item{} When $N \epsilon \ll 1$ the dynamics can be described by a
quantum adiabatic approximation. The gauge theory stays in the ground
state of the instantaneous Hamiltonian to good approximation.  At late
times the system is well described by smooth $AdS_5$ spacetime. This
is true even when $\lambda_{min}\le 1$ as discussed in \S4.

\item{} When $N\epsilon \gg 1$ and $\lambda_{min}\gg 1$, the system is well
described by a supergravity solution, which consists of $AdS_5$ spacetime
with corrections which are suppressed in $\epsilon$. The gauge theory provides
an alternate description in terms of weakly coupled harmonic oscillators which are
modes of gauge invariant operators dual to supergravity modes. These
oscillators  are
subjected to a driving force that is slowly varying compared to their
frequency. A classical adiabatic perturbation theory, the LNCAPT,
describes the dynamics of the system.  This dual description
reproduces the supergravity answers for the energy and $<\hat{{\cal O}}>$, as
discussed in \S6.1, \S6.2.

\item{} When $N\epsilon \gg 1$, and $\lambda_{min}\le O(1)$,
supergravity breaks down.  In this case we do not have a clean
conclusion for the final state of the system. Additional oscillators
which correspond to string modes can now get activated. There are two
possibilities :
either the
description in terms of classical adiabatic dynamics for the
oscillators continues to apply, or a qualitative new feature of
thermalisation sets in. In the former case spacetime in the far future
is well approximated by smooth AdS space.  In the latter case the
gravity description depends on the value of $\epsilon$ and may consist
of a string gas or small black holes. This is discussed
in \S6.3.

\item{} We have not addressed here what happens when the dilaton
begins to vary more rapidly and $\epsilon$ becomes $\sim O(1)$. It is
natural to speculate that a black hole forms eventually in this
case. The oscillators in the gauge theory now become strongly coupled
with $O(1)$ anharmonic couplings.

 If $\lambda_{min}\gg 1$ this parametric regime can be studied
in supergravity itself. When $\epsilon \ll 1$ the calculations in \S2 showed
that no black hole forms. As $\epsilon$ increases the natural
expectation is that eventually a black hole should begin to form at
some critical value. The size of this black hole should then grow with
$\epsilon$, leading to a big black hole with radius bigger than AdS
scale. Very preliminary indications for this come from the
calculations in \S2 where we see that as $\epsilon$ increases the
value of $|g_{tt}|$ becomes smaller at the center of AdS
eq.(\ref{feA}), suggesting that a horizon would eventually
form at $\epsilon \sim O(1)$. Better evidence comes from studying a
region of parameter space where $\epsilon \gg 1$ but where the total
amplitude of the dilaton variation is small. In this case
\footnote{The results reported in \cite{Bhattacharyya:2009uu} are for
the case of $AdS_{d+1}$ spacetimes with $d$ odd.} one finds that a
boundary variation of the dilaton, which is sufficiently fast compared
to its amplitude, always produces a black hole.

When $\lambda_{min}\le O(1)$, and $\epsilon$ becomes $\sim O(1)$,
supergravity breaks down at intermediate times. If thermalisation has already
set in in the parametric regime, $N \epsilon \gg 1, \epsilon \ll 1$,
as discussed above, then one expects that the small black hole which
has formed for $\epsilon \ll 1$ would grow and become of order the AdS
scale or bigger when $\epsilon \ge O(1)$. If thermalisation does not
set in when $\epsilon \ll 1$, then at some critical value $\epsilon
\sim O(1)$ one would expect that this does happen leading to the
formation of a black hole whose mass then grows as $\epsilon$ further
increases.

It will be interesting to try and analyse this regime further  in
subsequent work.

\item{} Finally one can consider a regime where $\epsilon \rightarrow
\infty$ at time $t\rightarrow 0$. This regime was considered in
\cite{adnt} where the dilaton was taken to vanishes like $e^\Phi\sim
(t)^p$ as $t\rightarrow 0,$ leading to a diverging value for
$\dot{\Phi}$. In a toy quantum mechanics model it was argued that the
response of the system in this case is singular, suggesting that this
singularity is a genuine pathology which is not smoothened
out. However the conclusions for the toy model do not directly apply
to the field theory. Important questions regarding the renormalisation
of this time dependent field theory remain and could invalidate this
conclusion.

\end{itemize}

One is hesitant to try and draw general conclusions about the
possibility of emergence of a smooth
spacetime from string scale curved regions on the basis
of the very limited analysis presented here. One lesson which has
emerged is that, at least for the kind of time dependence studied in
this paper, AdS space has a tendency to form a black hole
\footnote{AdS space is of course homogeneous so the reader might be
puzzled about where the black hole forms. The point is that the time
dependence imposed on the boundary picks out a particular notion of
time and the black hole forms where the redshift factor for this time
is smallest, this is the ``center of AdS space'' in global
coordinates. }. This fate can be avoided (as in the case when
$N\epsilon \ll 1$) but it requires slow time variation or perhaps more
generally rather finally tuned conditions. To understand in greater
detail when this fate of black hole formation can be avoided requires
a deeper understanding of the process of thermalisation in the dual
field theory.

In this paper we analysed the effects of a time dependent dilaton. It
will be interesting to extend this to other supergravity modes as well
by making their boundary values time dependent - e.g, making the
radius of the $S^3$ on which the gauge theory lives time dependent or
introducing time dependence along the other exactly flat directions in
the ${\cal N}=4$ theory besides the dilaton.  Also, we have kept the
parameter $N$ fixed in this work. As was discussed in the introduction
$N$ measures the strength of quantum corrections and is also the value
of $R_{AdS}$ in Planck units eq.(\ref{radiusb}).  It would be
interesting to consider cases where $N$ changes and become smaller
thereby increasing the strength of quantum effects and making the
curvature of order $l_{Pl}$. One way to do this might be by
introducing time dependence that moves the system onto the Coulomb
branch. This could reduce the effective value of $N$ in the interior.
For recent interesting work see, \cite{HLS}, also the related earlier
work, \cite{HH}, \cite{TCH}. Finally, a length scale was introduced in
the gauge theory  by working on $S^3$ here. Instead one could consider
a confining gauge theory like the Klebanov-Strassler kind \footnote{We thank M. Mulligan for a related
discussion.}, \cite{KS}, which has a mass gap on
$R^3$. In this case one could consider the response of the system to
time dependence slow compared to the confining scale and hope to use
an adiabatic approximation to understand this response.

\section{Acknowledgments}

We would like to thank  Ian Ellwood, Gary Horowitz, Shamit Kachru, Per Kraus, Steve
Shenker, Eva Silverstein, Spenta Wadia and especially Shiraz Minwalla
for discussions, and K. Narayan for discussions and collaboration at
the early stages of this work. The work of A.A. is partially
supported by ICTP grant Proj-30 and the Egyptian Academy
for Scientific Research and Technology.
 A.A. and A.G. would like to thank the
Chennai Mathematical Institute (and the organizers of Indian Strings
Meeting 2008) for hospitality. S.R.D. would like to thank the
International Center for Theoretical Sciences, Tata Institute of
Fundamental Research and the organizers of ``String Theory and
Fundamental Physics'' for hospitality. The work of S.R.D, A.G and J.O
is supported in part by a National Science Foundation (USA) grant
PHY-0555444.  S.T. thanks the organisers of the Monsoon Workshop,  at
TIFR,  for
the stimulating meeting during which some of this work was initiated.
S.T.  is on a sabbatical visit to Stanford University
and SLAC National Accelerator Laboratory for the period
Oct. 2008-Sept. 2009 and thanks his hosts for their kind hospitality
and support.
Most of all he thanks the people of
India for generously supporting research in string theory.

\appendix

\section{Comments on Metric to $O(\epsilon^2)$}
We are interested in calculating the back reaction on the metric to
$O(\epsilon^2)$ that arises due to the dilaton $\Phi_0$.  Without
loss of generality we can assume that the metric is $S^3$ symmetric
and therefore of form, \be
\label{fm1}
ds^2=-g_{tt}dt^2+g_{rr}dr^2+2g_{tr}dtdr + R^2 d\Omega^2 \ee where the
metric coefficients are functions of $r,t$.  The zeroth order metric
is that of $AdS_5$, eq.(\ref{metgc}). We argued above that the
backreaction to the dilaton source arises at order $\epsilon^2$. Thus
$g_{tr}$ in eq.(\ref{fm1}) is of order $\epsilon^2$.

We now show that by doing a suitable coordinate transformation, the
mixed component $g_{tr}$ can be set to vanish up to order
$\epsilon^2$.  The coordinate transformation is, from $(t,r)$ to
$(t,\tilde{r})$, where, \be
\label{ctaa}
r=\tilde{r}-{g_{tr}\over g_{rr}}t,
\ee
which leads to
\be
\label{ctab}
dr=d\tilde{r}-({g_{tr}\over g_{rr}})' t d\tilde{r}-{g_{tr}\over
g_{rr}}dt + O(\epsilon^3).  \ee Prime above indicates derivatives with
respect to $r$, We can drop the $\epsilon^3$ terms for our purpose,
these originate from additional time derivatives on the metric
components.  Substituting in eq.(\ref{fm1}) we see that in the new
coordinates the $g_{t\tilde{r}}$ components of the metric vanish upto
$O(\epsilon^3)$ corrections which we are neglecting anyways.  To avoid
clutter we will henceforth drop the tilde on the $r$ coordinate and
write the metric as \be
\label{fm2}
ds^2=-g_{tt}dt^2+g_{rr}dr^2+ R^2d\Omega^2
\ee

Next we show that up to $O(\epsilon^2)$ we can set $R$ equal to the
coordinate $r$ without reintroducing the mixed components.  First
define, \be
\label{defrp}
\bar{r}=R
\ee
leading to,
\be
\label{defrp2}
d\bar{r}=R'dr+\dot{R} dt \ee where dot indicates a time derivative.
Now any time dependence in $R$ arises only due to the dilaton and
therefore is of order $\epsilon^2$. This means that $\dot{R}$ is
$O(\epsilon^3)$ and can be neglected. So up to $O(\epsilon^2)$ no
mixed components arise in the metric due to this coordinate
transformation. We now drop the bar on the radial coordinate and write
the final metric as, \be
\label{fm3}
ds^2=-g_{tt}dt^2+g_{rr}dr^2+r^2 d\Omega^2.
\ee

\section{More  on the Driven Harmonic Oscillator}
\label{appb}
In this appendix we provide the steps leading to
(\ref{coherenteqns}) and (\ref{cstatepara}). The time derivative of
the state vector $|\psi(t)>$ in (\ref{statet}) is
\be
\label{schlhs}
i \frac{\partial}{\partial t} |\psi (t) > =
i(\dot{\lambda}+\dot{\alpha})a^\dagger |\psi(t)>+i({\dot{N}\over N}+{\dot{N_\alpha}
\over N_\alpha})|\psi(t)>
\ee
where we have used the expression for $|\phi_0>$ in
(\ref{agndstate}). The action of the hamiltonian $H$ on the state is
easily obtained by noting that
\ben
[ H , e^{\lambda a^\dagger} ] = \left( \omega_0 \lambda a^\dagger + \frac{J
  \lambda}{\sqrt{2\omega_0}} \right) e^{\lambda a^\dagger}.
\een
This leads to
\be
\label{schrhs}
H|\psi(t)>=\left(\omega_0\lambda a^\dagger+{J \lambda \over\sqrt{2\omega_0}}\right)
|\psi(t)>
+{\omega_0\over 2}|\psi(t)> .
\ee
It may easily be checked that the states $|\psi(t)>$ and $a^\dagger |\psi(t)>$ are linearly
independent. Equating the coefficients of $a^\dagger |\psi(t)>$
 in eq.(\ref{schlhs}) and (\ref{schrhs}) and using eq.(\ref{valpara})
 then leads to eq.(\ref{coherenteqns}). Equating the coefficients of $|\psi(t)>$
 in eq.(\ref{schlhs}) and (\ref{schrhs}) gives an equation
 that determines $N(t)$. Note that $|N(t)|$ is determined directly from the requirement
 that $<\psi|\psi>=1$.

\section{The normalization factor $F(2n)$} \label{normal}

In computing the normalization $F(2n)$ in (\ref{csix}) it is best to
first continue to euclidean signature and then perform a conformal
transformation from $R \times S^3$ to $R^4$. The radial coordinate on
the $R^4$ is given by $r = e^\tau$. where $\tau$ is the euclidean time
in $R \times S^3$. Then the Heisenberg picture operator on $R^4$ is
given by
\ben \hat{{\cal O}}_{l=0} = \sum_{m=-\infty}^\infty
\frac{\cO_{m}}{r^{m+4}}
\label{apone}
\een
The factor of $r^{m+4}$ in the denominator reflects the fact that the
operator $\hat{{\cal O}}_{l=0}$ has dimension 4. The conformally invariant
vacuum satisfies
\bea
\cO_m |0> & = & 0~~~~~~~m \geq -3 \nn \\
<0| \cO_m & = & 0~~~~~~~m \leq 3
\eea
Then the radial time ordered 2 point function is given by
\ben
< \hat{{\cal O}}_{l=0} (r) \hat{{\cal O}}_{l=0} (r^\prime) >
 =   \sum_{m=4}^\infty \sum_{n=-\infty}^{-4} \frac{<0|\cO_m \cO_n |0>}
{r^{m+4}(r^\prime)^{n+4}}
\label{aponee}
\een
The 2 point function only involves the central term in the operator
algebra. This means we can write
\bea
\cO_{m} & = & N F(m) A_{m}~~~~~~~~~(m > 0) \nn \\
\cO_{-m} & = & N  F^\star(m) A^\dagger_{m}~~~~~~~~~(m > 0)
\eea
where the operators $A_m, A^\dagger$ satisfies an operator algebra and
$F(m)$ is a normalization
\ben
[ A_m , A_n ] = [A^\dagger_m , A^\dagger_n ] = 0~~~~~~~~
[ A_m , A^\dagger_n ] = \delta_{mn}
\een
Note that because of (\ref{aponee}) only terms for $n \geq 4$ contribute to the sum.
This leads to the result
\ben
< \hat{{\cal O}}_{l=0} (r) \hat{{\cal O}}_{l=0} (r^\prime) >
 =  \frac{N^2}{r^8} \sum_{m=4}^\infty |F(m)|^2 \left( \frac{r^\prime}{r} \right)^{m-4}
\label{aptwo}
\een
On the other hand since the dimension of the operator $\hcOP (r,\Omega_3)$ is $4$ we know the 2 point function on $R^4$. This is given by
\ben
< \hat{{ \cal O}} (r,\Omega_3) \hat{{\cal O}} (r^\prime, \Omega^\prime_3)> = \frac{A N^2}{|{\vec r} - {\vec r}^\prime |^8}
\een
where $A$ is a order one numerical constant. Here ${\vec r} = (r, \Omega)$ etc., is  the location of the operator on $R^4$. Integrating over $\Omega_3, \Omega_3^\prime$ we get
\ben
\int d\Omega_3 \int d\Omega_3^\prime < \hat{{\cal O}} (r,\Omega_3)
\hat{{\cal O}} (r^\prime, \Omega^\prime_3)> = A N^2(8 \pi^3) \int_0^\pi \frac{\sin^2 \theta~d\theta}{(r^2 + (r^\prime)^2 - 2 r r^\prime \cos \theta)^4}
\een
The integral can be performed. The result is, for $r > r^\prime$
\ben
\int d\Omega_3 \int d\Omega_3^\prime < \hat{{\cal O}} (r,\Omega_3)
\hat{{\cal O}} (r^\prime, \Omega^\prime_3)> = N^2 \frac{4A \pi^4}{r^8} \frac{ \left( \frac{r^\prime}{r} \right)^2 + 1}{( 1- \left( \frac{r^\prime}{r} \right)^2)^5}
\een
Using the power series expansion
\ben
\frac{1+x}{(1-x)^5} = \sum_{m=0}^\infty \frac{1}{12} (m+1)(m+2)^2(m+3) x^m
\een
we finally get
\ben
\int d\Omega_3 \int d\Omega_3^\prime < \hat{{\cal O}} (r,\Omega_3)
\hat{{\cal O}} (r^\prime, \Omega^\prime_3)> = N^2 \frac{A \pi^4}{3}\frac{1}{r^8}\sum_{m=0}^\infty
(m+1)(m+2)^2(m+3)  \left( \frac{r^\prime}{r} \right)^{2m}
\label{apthree}
\een
The result clearly shows that only operators with even mode numbers are present in the expansion (\ref{apone}). Comparing (\ref{apthree}) and
(\ref{aptwo}) we get
\ben
F(2m+1) = 0~~~~~~~~~~~~
|F(2m)|^2  =  \frac{A\pi^4}{3} m^2 (m^2-1)
\een
which is the result in equation (\ref{csix}).

\end{document}